\documentclass[prd,nofootinbib,preprint]{revtex4}
\usepackage{graphicx}
\usepackage{epsfig}
\usepackage{amsmath}
\usepackage{amsfonts}
\usepackage{amssymb}
\usepackage{url}
\usepackage{hyperref}
\usepackage{subfigure}
\usepackage{hhline}
\usepackage{color}
\usepackage{bm}
\usepackage{cancel}
\newcommand{\beqa}{\begin{eqnarray}}
\newcommand{\eeqa}{\end{eqnarray}}
\newcommand{\beq}{\begin{equation}}
\newcommand{\eeq}{\end{equation}}

\newcommand{\nn}{\nonumber}
\newcommand{\bmt}{\begin{pmatrix}}
\newcommand{\emt}{\end{pmatrix}}
\usepackage[toc,page]{appendix}
\usepackage{comment}
\usepackage{hyperref}
\newcommand{\be}{\begin{equation}}
\newcommand{\ee}{\end{equation}}
\newcommand{\bea}{\begin{eqnarray}}
\newcommand{\eea}{\end{eqnarray}}

\begin{document}
\title{Light sterile neutrinos and their implications on  currently running long-baseline  and neutrinoless double beta decay experiments
}
\author{Rudra Majhi$^{1}$}
\email{rudra.majhi95@gmail.com}

\author{Soumya C.$^{2}$}
\email{soumyac20@gmail.com}

\author{Rukmani Mohanta$^{1}$}
\email{rmsp@uohyd.ac.in}

\affiliation{$^1$\,School of Physics, University of Hyderabad,
              Hyderabad - 500046, India \\                   
 $^2$\,Institute of Physics, Sachivalaya Marg,   Sainik School Post, Bhubaneswar 751005, India. }

\begin{abstract}
Recent $\nu_e$ appearance data from the Mini Booster Neutrino Experiment (MiniBooNE) are in support of  the excess of events reported by the Liquid Scintillator Neutrino Detector (LSND), which provides an indirect hint for the existence of eV-scale sterile neutrino.  As these sterile neutrinos can mix with the standard active neutrinos, in this paper we explore the effect of such active-sterile mixing on the determination of various oscillation parameters by the currently running long-baseline neutrino experiments T2K and NO$\nu$A. We find that the existence of sterile neutrino can lead to new kind of degeneracies among these parameters which would substantially deteriorate the mass hierarchy  sensitivity of NO$\nu$A experiment. We further notice that the inclusion of   data from  T2K experiment helps in resolving  the degeneracies.
The impact of new CP violating phases  $\delta_{14}$ and $\delta_{34}$ on the maximal CP-violation exclusion sensitivity for NO$\nu$A experiment has also been illustrated. Finally, we discuss the implication of such light sterile neutrinos on neutrinoless double beta decay processes in line with recent experimental results, as well as on the sensitivity reach of future experiments.
\end{abstract}
\maketitle
 
 \section{Introduction}

Unlike other fundamental fermions, neutrinos posses several unique features and are considered to be massless in the Standard Model (SM). However, the experimental observation of neutrino oscillation by various experiments, wherein the neutrinos change their flavour as they propagate,  provides  compelling evidence that at least two out of the three   neutrinos in the SM will have tiny but non-zero masses. In this regard, tremendous attempts are being made to understand the origin of  their masses, mixing phenomena, mass scale, whether they are  Dirac or  Majorana type in nature, etc. The three-flavour oscillation picture can successfully explain the  experimental results from solar, atmospheric and reactor neutrino experiments \cite{Tanabashi:2018oca}. In this framework, the phenomenon of neutrino oscillation is characterized by three mixing angles ($\theta_{12},\theta_{13},\theta_{23}$),
 two mass squared differences $\Delta m^2_{21}, ~\Delta m^2_{31}$ and one Dirac type CP phase $\delta_{\rm CP}$. These  oscillation parameters are measured very precisely, though there are a few unknowns, which are yet to be determined, like the neutrino mass ordering, octant of the atmospheric mixing angle $\theta_{23}$ and the CP violating phase $\delta_{\rm CP}$.
The main physics goal of current and future generation neutrino oscillation experiments is to precisely determine all these unknowns.  In this context,  the long-baseline experiments play a crucial role in the determination of these  parameters \cite{Agarwalla:2014fva, Feldman:2013vca}, due to the presence of enhanced matter effect. However, the existence of parameter degeneracies among the oscillation parameters greatly affect the sensitivities of these experiments \cite{Barger:2001yr}. Therefore, the resolution of parameter degeneracies  is the primary concern in neutrino oscillation studies.
 
 Apart from these, another important aspect in the neutrino sector is the possible existence  of additional eV-scale sterile neutrino  species ($\nu_s$), which has attracted a lot of attention in recent times,  following some anomalies reported by various experiments.  The first such anomaly was presented by the Liquid Scintillator Neutrino Detector (LSND) experiment \cite{Aguilar:2001ty},  in the measurement of anti-neutrino flux in $\overline{ \nu}_\mu \to \overline{ \nu}_e$ oscillation.  An excess in the electron anti-neutrino ($\overline{\nu}_e$) events has been reported, which  could be explained by incorporating at least one additional eV-scale  neutrino. This result was further supported by the   $\overline \nu_e$ appearance results at the MiniBooNE experiment \cite{Aguilar-Arevalo:2013pmq}.  Another hint for  existence of light sterile neutrinos has emerged from the deficit in  the estimated anti-neutrino flux from reactor experiments \cite{Lasserre:2012vy, Mention:2011rk}.   Recent  measurements of the ratios of inverse beta-decay energy spectra by the short-baseline experiments NEOS \cite{Ko:2016owz} and DANSS \cite{Alekseev:2018efk}, at different distances also  appear to  exhibit some preference for sterile neutrino oscillations, while other recent short-baseline measurements, PROSPECT \cite{Ashenfelter:2018iov} and STEREO \cite{Lhuillier:2015fga}, don't show any such evidence.  Similar anomalies have also been observed at  GALLEX \cite{Anselmann:1994ar, Hampel:1997fc,Kaether:2010ag} and SAGE \cite{Abdurashitov:1996dp} Gallium experiments for solar neutrino observation,  which   indicate the existence of additional light neutrino species \cite{Giunti:2012tn,Giunti:2010zu, Abazajian:2012ys, Abdurashitov:2009tn, Hampel:1997fc}.
 Recently  MiniBooNE collaboration  \cite{Aguilar-Arevalo:2018gpe}, reported their new analysis with twice the  data sample size used earlier, confirming the anomaly at the level of 4.8$\sigma$, which becomes $>6\sigma$, if combined with LSND data.
 However, no evidence of active-sterile neutrino mixing  is observed  by the MINOS and MINOS+ \cite{Adamson:2017uda}, and  the joint analysis of these experiments sets stringent limits on the active-sterile mixing angles for values  $\Delta m_{41}^2 >10^{-2}~{\rm eV}^2$, through the study of $\nu_\mu$ disappearance. 
More importantly,  the entire MiniBooNE 90\% C.L. allowed  region is excluded by  MINOS/MINOS+ at 90\% C.L..  This in turn  implies a tension between MiniBooNE and MINOS/MINOS+ results.
 Recently, NO$\nu$A \cite{NOvA:2018gge}  has also performed the search for active-sterile neutrino mixing using neutral current interactions, though no evidence of $\nu_\mu \to \nu_s$ has been found. 
In this direction, Fermilab's  short-baseline neutrino program, the  ICARUS experiment, as well as the SoLid experiment at Belgium are expected  to provide definitive answer to  the long-standing anomalous results related to sterile neutrinos.

Though  the possible existence of an eV-scale neutrino could  explain  the above mentioned reactor as well as the LSND and MiniBooNE anomalies, it cannot have gauge interactions with the SM gauge bosons,  thus  safeguarding  precision measurement of $Z$ boson decay width by LEP experiment. 
 Hence, such a neutrino is known as  sterile neutrino, while the usual standard model neutrinos are known as active neutrinos.  Though sterile neutrinos are blind to weak interactions,  they can mix with active neutrinos. Therefore, in this paper we explore the effect of such active-sterile mixing on the determination of neutrino oscillation parameters  by currently running long-baseline neutrino experiments.   We, further investigate  its effect on neutrinoless double beta decay process. The implications  of light sterile neutrino on the physics potential of various long-baseline experiments, such as T2K, T2HK, NO$\nu$A and DUNE have been explored by several authors \cite {Dutta:2016glq,Ska:2018,Gupta:2018qsv,Ghosh:2017atj,Agarwalla:2016mrc, Palazzo:2015gja, Dewhurst:2015aba, Bhattacharya:2011ee, Chatla:2018sos, Choubey:2017cba, Agarwalla:2016xxa, Berryman:2015nua,Choubey:2017ppj, Gandhi:2015xza} for various possible combinations of run-period.  However, in this work we focus our attention on the following aspects. First, 
 we would like to see   whether the determination of mass-ordering by the currently running long-baseline experiments NO$\nu$A and T2K would be affected by the presence of light sterile  neutrinos.  Next, as the recent global fit hints towards the possibility of maximal CP violation in the neutrino sector, i.e., $\delta_{\rm CP} \approx  3\pi/2$ \cite{deSalas:2017kay},  we therefore, investigate the sensitivity of these experiments for the exclusion of maximal CP-violation scenario. We also briefly demonstrate the implications of light sterile neutrinos on neutrinoless double beta decay. A part of this work has been reported in \cite{Majhi:2019exu}. The impact of one sterile neutrino in the sensitivity studies of these two experiments towards neutrino mass hierarchy  and  CP violation discovery has been performed in a recent work \cite{Agarwalla:2016mrc}. However, our work differs from their analysis in the following ways. Firstly, we have considered all the three new mixing angles as well as the additional two CP violating phases  to be non-zero, whereas they have assumed $\sin^2\theta_{34}=0$ and $\delta_{34}=0$. Secondly, concerning the CP violation studies, we have performed the analysis for the exclusion of maximal CP-violation scenario,  which has been done for the first time to the best of our knowledge. 
  
The paper is organised as follows.  In section II, we discuss the theoretical framework for (3+1) flavor oscillation scheme. Section III covers the experimental set-up and details about the analysis adopted in this paper. The effect of light sterile-neutrino on oscillation parameters  is discussed in section IV. Section V deals with mass hierarchy (MH) sensitivity, while the maximal CP violation sensitivity  is discussed  in Section VI. Section VII focused on the impact of sterile neutrino on neutrinoless double beta decay, prior to conclusion in section VIII.

\section{Brief Discussion about 3+1 Oscillation Model}
In the presence of one sterile neutrino, the so-called 3+1 scenario, there will be mixing between the three active neutrinos with the sterile one and hence, the neutrino mixing matrix can be represented by a  $4\times 4$ unitary matrix.  Consequently, the parametrization of the neutrino mixing matrix requires some additional parameters, which includes
three  mixing angles ($\theta_{14},\theta_{24},\theta_{34}$) and two phases ($\delta_{14},\delta_{34}$).  Thus, analogous to the standard PMNS matrix, the four dimensional mixing matrix will have the form
\begin{equation}
U^{3+1}= O(\theta_{34},\delta_{34}) R(\theta_{24}) O(\theta_{14},\delta_{14})R(\theta_{23}) O(\theta_{13},\delta_{13}) R(\theta_{12})\;,
\end{equation} 
where $R(\theta_{ij})$ ($O(\theta_{ij},\delta_{ij})$) are the real (complex)  $4\times 4$ rotation matrices  in the $ij$ plane,   which contain the  $2\times 2$ sub-matrices
\begin{equation}
R^{2 \times 2} (\theta_{ij})= \left( \begin{array}{cc}
\cos \theta_{ij}& \sin\theta_{ij} \\
-\sin \theta_{ij} & \cos\theta_{ij}
\end{array}
\right),~~~~~
O^{2 \times 2} (\theta_{ij},\delta_{ij})= \left( \begin{array}{cc}
\cos \theta_{ij}& \sin\theta_{ij} e^{-i\delta_{ij}}\\
-\sin \theta_{ij}e^{i\delta_{ij}} & \cos\theta_{ij}
\end{array}
\right),
\end{equation}
 as the $ij$ sub-block. 
 Incorporating the $4\times 4$ mixing matrix (1), the oscillation probability for $\nu_\mu\to \nu_e$ transition in the 3+1 framework can be expressed  in terms of the effective mixing matrix elements ($\tilde{U}_{\alpha i}$) and effective mass square differences $\Delta \tilde{m}^2_{ij}$, in presence of matter  as \cite{Li:2018ezt}
\begin{eqnarray} 
&&P(\nu_\mu\to \nu_e)=\sum_i |\tilde{U}_{\mu i}|^2 |\tilde{U}_{e i}|^2+
 2\sum_{i<j}\left[{\rm Re}(\tilde{U}_{\mu i}\tilde{U}_{e j}\tilde{U}^* _{\mu j}\tilde{U}^* _{e i})\cos \Delta_{ij}-{\rm Im}(\tilde{U}_{\mu i}\tilde{U}_{e j}\tilde{U}^* _{\mu j}\tilde{U}^* _{e i})\sin \Delta _{ij}\right],\nn\\
\end{eqnarray}
where  $\Delta _{ij}=\Delta \tilde{m}^2_{ij}L/2E$, $L$ and $E$ are the baseline and  energy of neutrino beam, respectively. The effective mass squared difference $\Delta \tilde{m}^2_{ij}$ can be written in terms of two arbitrary  mass squared differences as
\begin{eqnarray}
&&\Delta \tilde{m}^2_{ij}=\hat{\Delta} m^2_{i1}-\hat{\Delta} m^2_{j1}\;.
\end{eqnarray}
The exact analytical expressions for $\hat{\Delta} m^2_{i 1}(i = 1,2,3,4)$ can be found in \cite{Li:2018ezt}. The effective mixing elements can be related  to the $4\times4$ mixing matrix elements ($U_{\alpha i}$, $\alpha=e,\mu,\tau,s$) as
\begin{eqnarray}
\tilde{U}_{e i}\tilde{U}_{\mu j}^*=\frac{1}{\prod\limits_{k\neq i}\Delta \tilde{m}^2_{ik}}\left[\sum_j F_{e\mu}^{ij}U_{ej}U^*_{\mu j}+C_{e\mu}\right]\;,
\eea
where
\bea
F_{e\mu}^{ij}&=&A^2\Delta m^2_{j1}+A\Delta m^2_{j1}(\Delta m^2_{j1}-\sum_{k\neq i}\hat{\Delta}m^2_{k1})+(\Delta m^2_{j1})^3-\sum_{k\neq i}(\Delta m^2_{j1})^2\hat{\Delta}m^2_{k1} \nn\\
&&+\sum_{k,l;k\neq l\neq i}\Delta m^2_{j1}\hat{\Delta}m^2_{k1}\hat{\Delta}m^2_{l1}\;,\nn\\
C_{e\mu}&=& A^\prime \sum_{kl}\Delta m^2_{k1}\Delta m^2_{l1}U_{ek}U_{\mu l}^*U_{sk}U_{sl}^*+A\sum_{k,l}\Delta m^2_{k1}\Delta m^2_{l1}|U_{ek}|^2U_{el}U_{\mu l}^*\;,
\end{eqnarray} 
with $A=2\sqrt{2}G_F N_eE,~A^\prime=-\sqrt{2}G_F N_nE$ and $N_e(N_n)$ is the electron (neutron) density. One can get back the oscillation  probability for three  neutrino scenario  in the presence of matter,  from the 3+1 case by assuming $U_{\alpha 4}=0,~ U_{si}=0$, and $ A^\prime =0$.

\begin{figure}[!htb]
\includegraphics[scale=0.285]{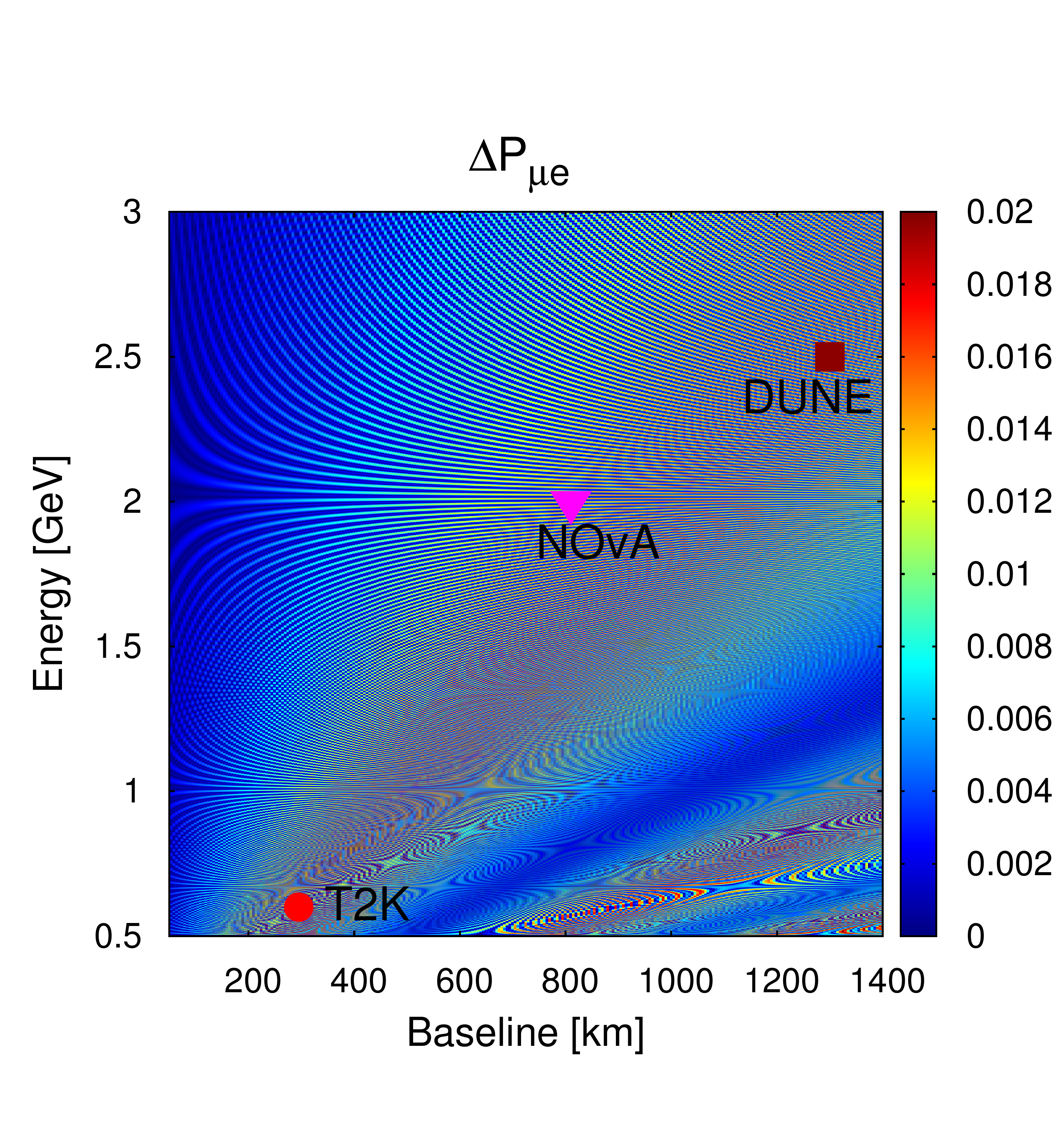}
\hspace*{0.1 true cm}
\includegraphics[scale=0.285]{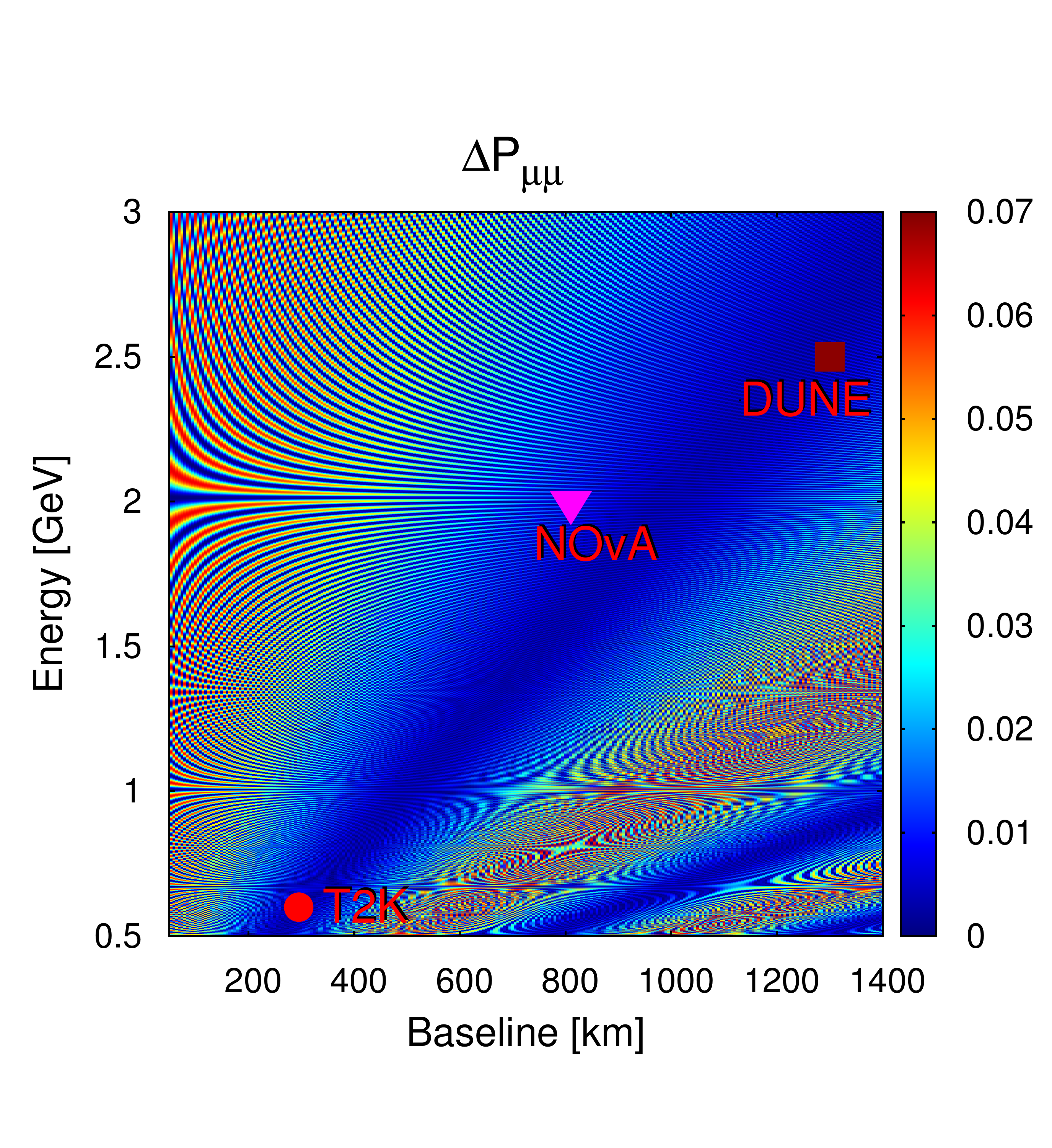}
\caption{ Graphical representation of $\Delta P_{\mu e}~ (\Delta P_{\mu \mu})$  in  the $L-E$ plane in left (right) panel. }
\label{oscillogram}
\end{figure}
Now, we would like to see the impact of a sterile neutrino on the physics potentials of accelerator based long-baseline  neutrino oscillation experiments, which are primarily designed to study  $\nu_\mu \to \nu_e$ and $\overline{\nu}_\mu \to \overline{\nu}_e$ oscillation channels. The first and foremost  implication  can be demonstrated by defining a quantity $\Delta P_{\alpha \beta}$, which is the absolute difference between the oscillation probabilities   in the presence of a sterile neutrino and the standard three flavor interaction (SI) scenario  in  presence of matter, i.e.,   $\big (\Delta P_{\alpha \beta}=| P_{\alpha\beta}^{\rm  sterile}-P_{\alpha\beta}^{\rm SI}|\big )$. Analogously, one can also construe the corresponding parameter for  anti-neutrino case  as $\overline{\Delta P}_{\alpha\beta}$. 
In  Fig.-\ref{oscillogram}, we show the graphical representation of  oscillograms  for $\Delta P_{\mu e}~(\Delta P_{\mu \mu})$ in left (right) panel, as function of baseline ($L$) and energy ($E$) for neutrino beam.  For obtaining these oscillograms, 
we have used the best-fit oscillation parameters as given in the Table  \ref{table}. In these plots, dark red regions represent  large deviation between the oscillation probabilities.  Moreover, it is clear from $\Delta P_{\mu e}$  plot that one can probe the effect of sterile neutrinos  in the  long-baseline experiments like T2K  ($L=295$ km and $E$=0.6 GeV), NO$\nu$A ($L$=810 km, $E$=2 GeV) and DUNE ($L$= 1300 km, $E= 2.5~ {\rm GeV}$). Hence, sterile neutrinos may play a crucial role in the determination of the oscillation parameters  in long-baseline neutrino oscillation experiments. Similarly, the sensitivity of $\Delta P_{\mu \mu}$   towards the presence of sterile neutrino for these experiments is also non-negligible ($\sim 1\%$),  as seen from the right panel of Fig. 1. 
\begin{table} 
\begin{tabular}{|c|c|c|} \hline
Parameters            & True values               & Test value Range  \\ \hline
$\sin^2 \theta_{12}$  & 0.310 & NA      \\ 
$\sin^2 \theta_{13}$ & 0.0224                   & NA \\ 
$\sin^2 \theta_{23} $ & 0.58                       & $0.4\rightarrow 0.62$\\& (LO 0.42)&$0.4\rightarrow 0.5$\\& (HO 0.58)&$0.5\rightarrow 0.62$ \\ 
$\delta_{\rm CP} $       & $ -90^\circ$                  & $-180^\circ \rightarrow 180^\circ $\\ 
$\Delta m^2_{21}$    & $7.39 \times 10^{-5}~{\rm eV}^2 $ & 
NA \\ 

$\Delta m^2_{31}$    & $ +2.525 \times 10 ^{-3}~{\rm  eV}^2$ (NH)& $+\textit{•}(2.43 \rightarrow 2.63) \times 10^{-3} {\rm eV}^2$ \\
& $ -2.512 \times 10 ^{-3}~{\rm eV}^2$ (IH)& $(-2.61 \rightarrow -2.41) \times 10^{-3}~{\rm eV}^2$ \\ 

$\Delta m^2_{14}$ & $1~{\rm eV}^2 $ & NA \\ 

$\sin^2 \theta_{14}$ & 0.0204 & ($0.0098 \to 0.031$) \\

$\sin^2 \theta_{24}$ & 0.0163 & ($0.006\to 0.0268$) \\

$\sin^2 \theta_{34}$ & 0.0197 & ($0 \to 0.0413$)\\ 
 
$\delta_{14} $ & $-90^\circ$ & $-180^\circ \rightarrow 180^\circ$\\ 
$\delta_{34} $ & $-90^\circ $ & $ -180^\circ \rightarrow 180^\circ$\\ \hline
\end{tabular}
\caption{ Values of oscillation parameters considered in our analysis are taken from the latest NuFIT results \cite{Esteban:2018azc}. Values for the sterile mixing angles and their allowed ranges  are calculated from the 3$\sigma$  ranges of the matrix elements $|U_{\alpha 4}|$ as discussed in \cite{Gariazzo:2017fdh}.}
\label{table}
\end{table}
\section{Simulation details}
As we are interested in exploring the impact of an eV-scale sterile neutrino on currently running long baseline experiments NO$\nu$A and T2K, we simulate these experiments using GLoBES software package  along with snu plugin \cite{Huber:2004gg,Huber:2009cw}. The auxiliary files and experimental specification of these experiments that we use in our analysis are taken from \cite{C.:2014ika}.  T2K and NO$\nu$A are complementary accelerator-based experiments with similar capabilities and goals, but differ only on their baselines. NO$\nu$A experiment is optimised to study the appearance of $\nu_e (\overline{\nu}_e)$ from a beam of $\nu_\mu (\overline{\nu}_\mu)$, consists of two functionally identical detectors, each located  14.6 mrad off the central-axis of Fermilab's neutrino beam, to receive a narrow band neutrino beam  with peak energy near 2 GeV,  corresponding to $\nu_\mu \to \nu_e$ oscillation maximum.   Its  near detector  of  mass 280 ton is located   about 1 km downstream (100 m  underground)  from the  source to measure un-oscillated  beam of muon-neutrinos and estimate backgrounds at the far detector.  Oscillated neutrino beam is observed by 14 kton far detector, situated in Ash  River,  810 km away from Fermilab. In order to do the simulation for  NO$\nu$A experiment, we consider  120 GeV proton beam energy with $6\times 10^{20}$ POT per year. We assume  signal efficiencies for both electron (muon) neutrino and anti-neutrino as 45\% (100\%). The background efficiencies for mis-identified muons (anti-muons) at the detector are considered as  0.83\% (0.22\%). The neutral current background efficiency for $\nu_\mu$ ($\overline{ \nu}_\mu$) is  assumed to be 2\% (3\%).  We further assume the intrinsic beam contamination, i.e., the background contribution coming from the existence of electron neutrino (anti-neutrino) in the beam to be about 26\% (18\%). Apart from these, we also consider   5\% uncertainty on signal normalization and 10\% on background normalization.  

The muon neutrino beam of T2K experiment is produced at Tokai and  is directed towards the water Cherenkov detector of fiducial mass  22.5 kton kept 295 km far away at Kamioka \cite{Abe:2017uxa}. The neutrino flux peaks  around 0.6 GeV as the detector is kept $2.5^{\circ}$ off-axial to the neutrino beam direction. In order to simulate T2K experiment, we consider the proton beam power of 750 kW and with  proton energy of 30 GeV, which corresponds to a total exposure of 7.8 $\times 10^{21}$ protons on target (POT) with 1:1 ratio of neutrino to anti-neutrino modes. We match the signal and back-ground event spectra and rates as given in the recent publication of the T2K collaboration \cite{Abe:2014tzr}. We consider an uncorrelated 5\% normalization error on signal and 10\% normalization error on background for both the appearance and disappearance channels as given in Ref.  \cite{Abe:2014tzr} to analyse the prospective data from the T2K experiment. We assume that the set of systematics for both the neutrino and anti-neutrino channels are uncorrelated.

We simulate the true ($N^\textrm{true}$) and test ($N^\textrm{test}$) event rates and compare them by using binned $\chi^2$ method defined in GLoBES, i.e.,
\begin{equation}
\chi^2_{\rm stat} (\vec{p}_\textrm{true},\vec{p}_\textrm{test})= \sum_{i\in\textrm{bins}} 2\Biggr[N_i^\textrm{test} - N_i^\textrm{true}-N_i^\textrm{true} \ln\left(\frac{N_i^\textrm{test}}{N_i^\textrm{true}}\right) \Biggr],
\end{equation}
where $\vec{p}$ stands for the array of standard neutrino oscillation parameters. However, for numerical evaluation of  $\chi^2$, we also incorporate the systematic errors using pull method, which is  generally done with the help of nuisance  parameters as discussed in the GLoBES manual.  Suppose $\vec{q}$ denotes the oscillation parameter in presence of sterile neutrino, then the  Mass Hierarchy (MH) sensitivity is given by 
\begin{eqnarray}
\chi^2_\text{MH}(\vec{q}) &=& \chi^2_\text{IH}(\vec{q}) -\chi^2_\text{NH}(\vec{q})~~~~(\text{for true Normal Hierarchy (NH)})\\
 \chi^2_\text{MH}(\vec{q}) &=& \chi^2_\text{NH}(\vec{q}) -\chi^2_\text{IH}(\vec{q})~~~~(\text{for true Inverted Hierarchy (IH)})
\end{eqnarray}
Further, we obtain minimum  $\chi^2_\text{min}$ by doing marginalization over all oscillation parameter spaces. In our analysis, we do not explicitly simulate the near detector for these experiments which may
provide some information about the active-sterile  mixing angles  $\theta_{i4}$,  but certainly, the near detector data is blind
to the CP phases, whose implications are mainly explored in this work. 
\section{Degeneracies  among oscillation parameters}
In this section, we discuss the degeneracies among the oscillation parameters in  presence of an eV-scale sterile neutrino. Here, we focus only on NO$\nu$A experiment.
  \begin{figure}[!htb]
\includegraphics[scale=1.0]{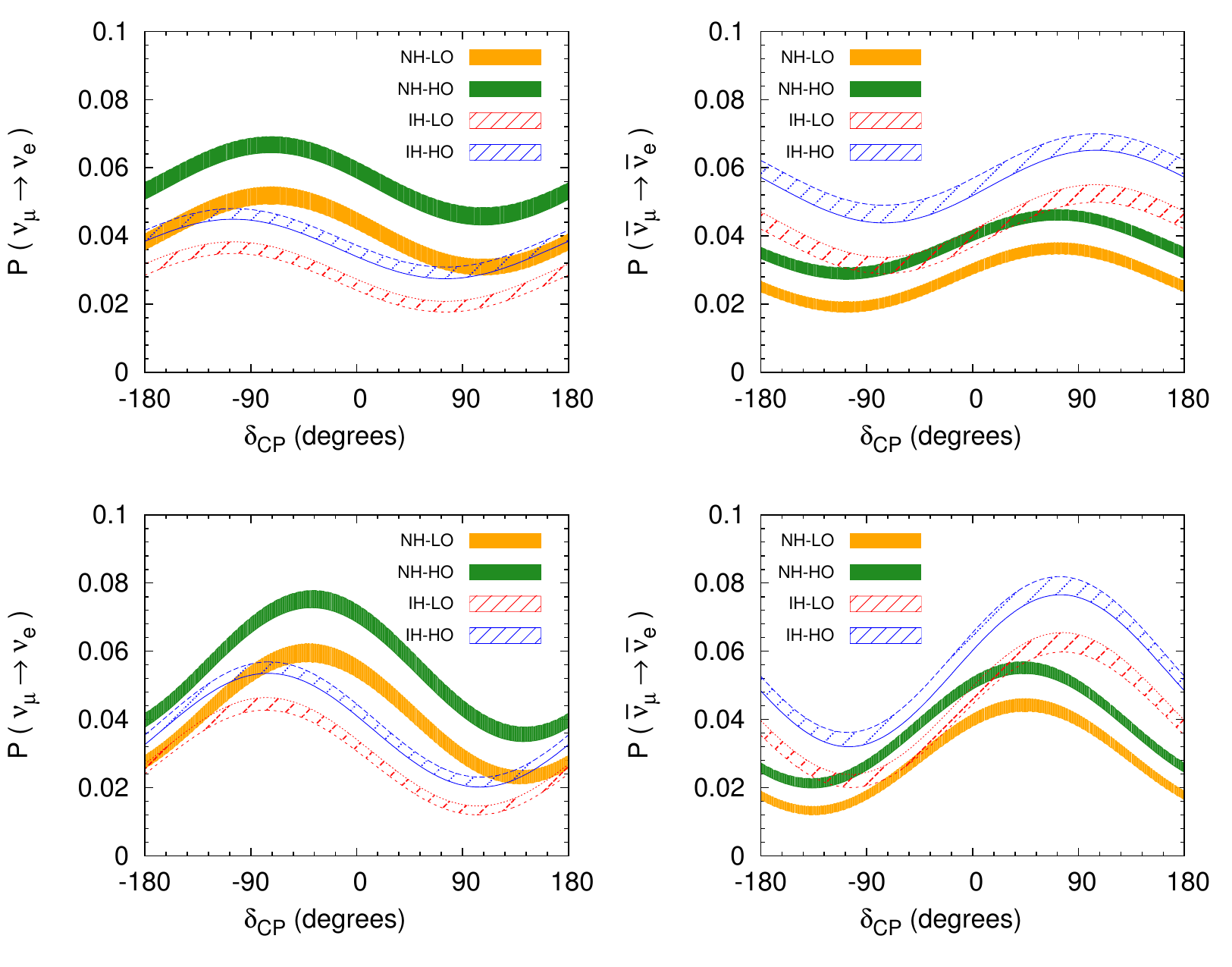}
\caption{ The  neutrino (anti-neutrino) oscillation probability as a function of $\delta_{\rm CP}$ is shown in the left (right) panel. The upper panel is for 3-flavor case, while the lower panel is for 3+1 case with $\delta_{14}=-90^\circ $ and $\delta_{34}=-90^\circ$.}
\label{probability}
\end{figure}

 In order to analyse degeneracies among the oscillation parameters at probability level, we show $\nu_e$ ($\overline{\nu}_e$) appearance  probability as a function of $\delta_{\rm CP}$  in the left (right) panel of Fig. \ref{probability}. The upper panel of the figure corresponds to oscillation probability in standard paradigm and that for 3+1 case is given in lower panel. The green, orange, blue and red bands in  the figure represent the oscillation probabilities for possible hierarchy-octant combinations:  NH-HO, NH-LO, IH-HO and IH-LO respectively, where HO and LO stand for higher octant and lower octant of $\theta_{23}$. From the upper panel of the figure, it can be seen that the bands for NH-HO and IH-LO  are very well separated in neutrino channel, whereas the  NH-LO and IH-HO bands are overlapped with each other, which results  degeneracies among the oscillation parameters. Also it should  be noted that, in the anti-neutrino channel, the case is just opposite. Therefore, a combined analysis of neutrino and anti-neutrino data helps in the resolution of degeneracies and also improves the sensitivity of long-baseline experiments to precisely determine  the unknowns of  standard oscillation paradigm.  From the bottom panel of the figure, it can be seen that  new types of degeneracies among the oscillation parameters have emerged, in the presence of sterile neutrino even for  a single value of the new CP phases $\delta_{14}$ ($=-90^\circ$) and $\delta_{34}$ ($=-90^\circ$), which can worsen the sensitivity of the unknowns. 
\begin{figure}[!htb]
\includegraphics[scale=0.55]{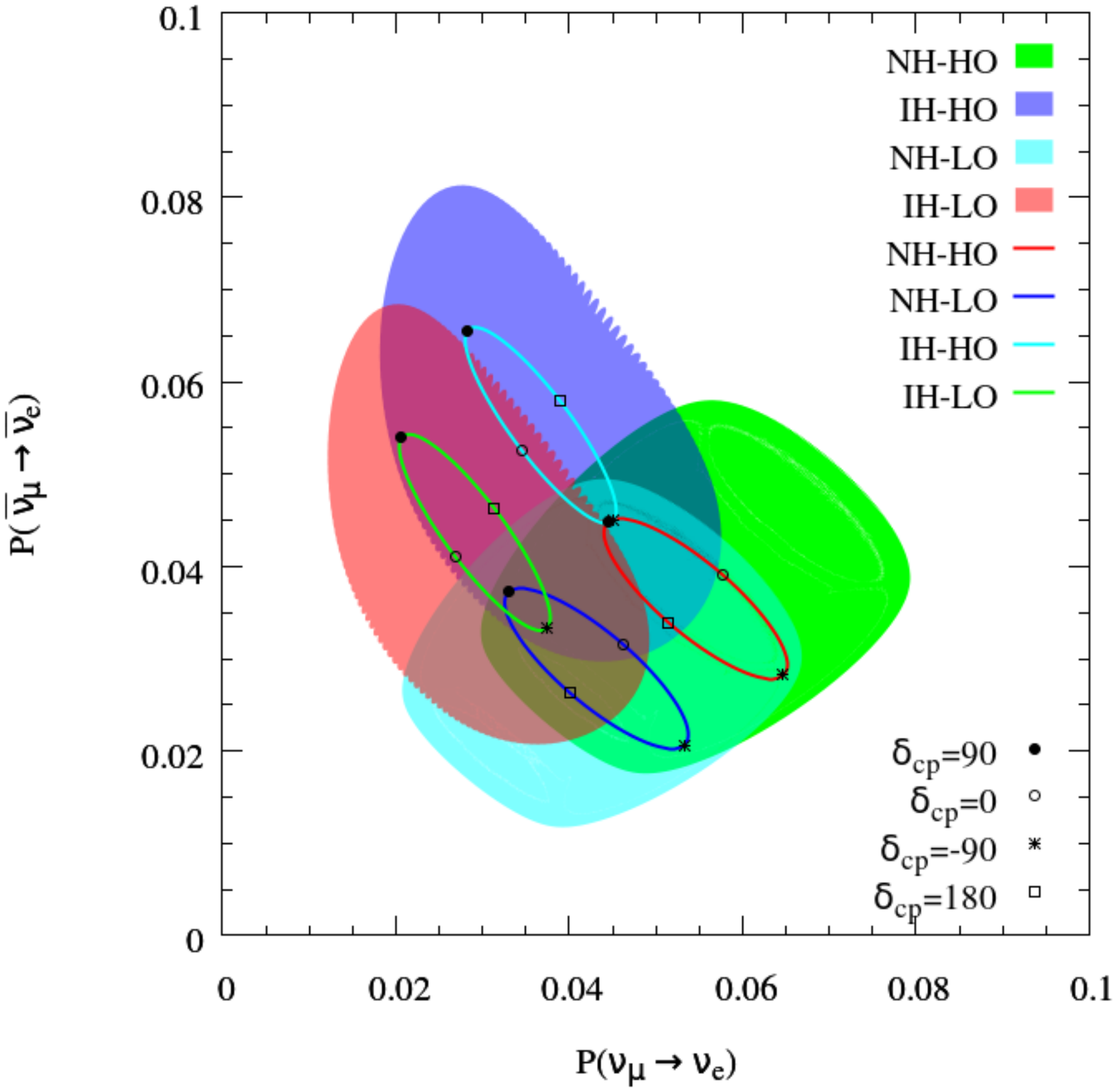}
\caption{Bi-probability plots for NO$\nu$A in 3 years in neutrino and 3 years in anti-neutrino mode for different hierarchy and octant combinations. }
\label{biprob}
\end{figure}

\begin{figure}
\begin{center}
\includegraphics[width=0.49\linewidth]{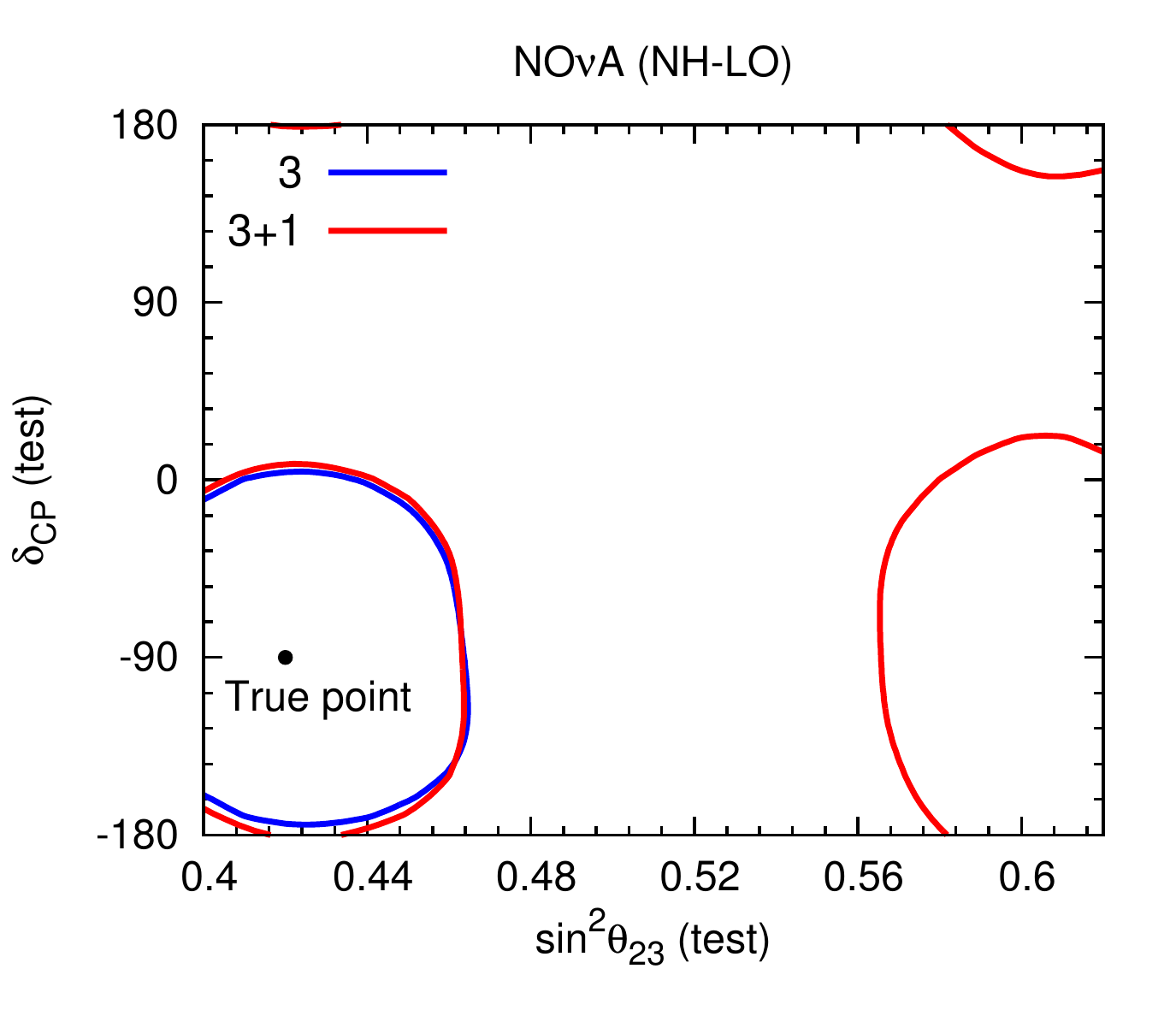}
\includegraphics[width=0.49\linewidth]{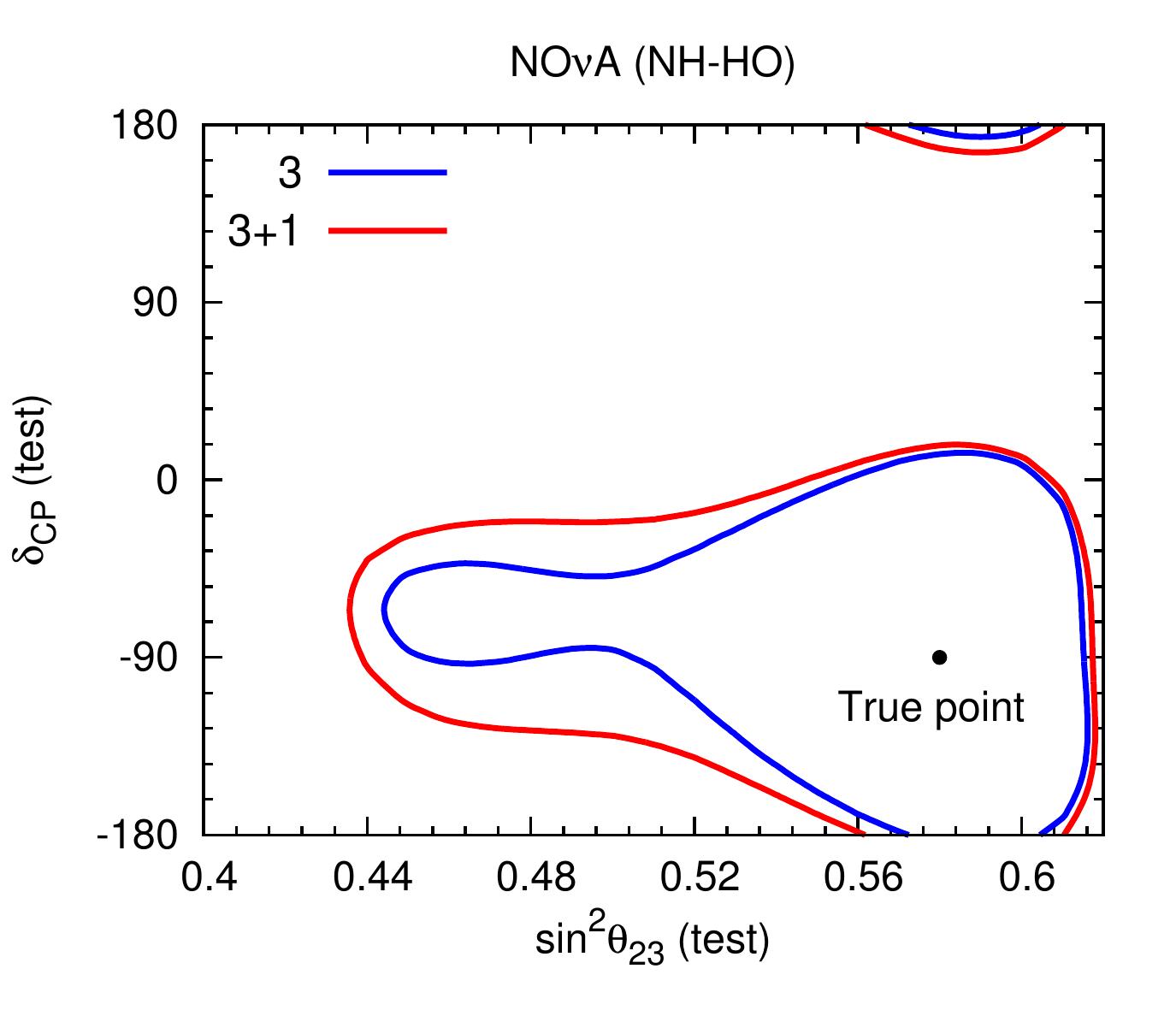}\\
\includegraphics[width=0.49\linewidth]{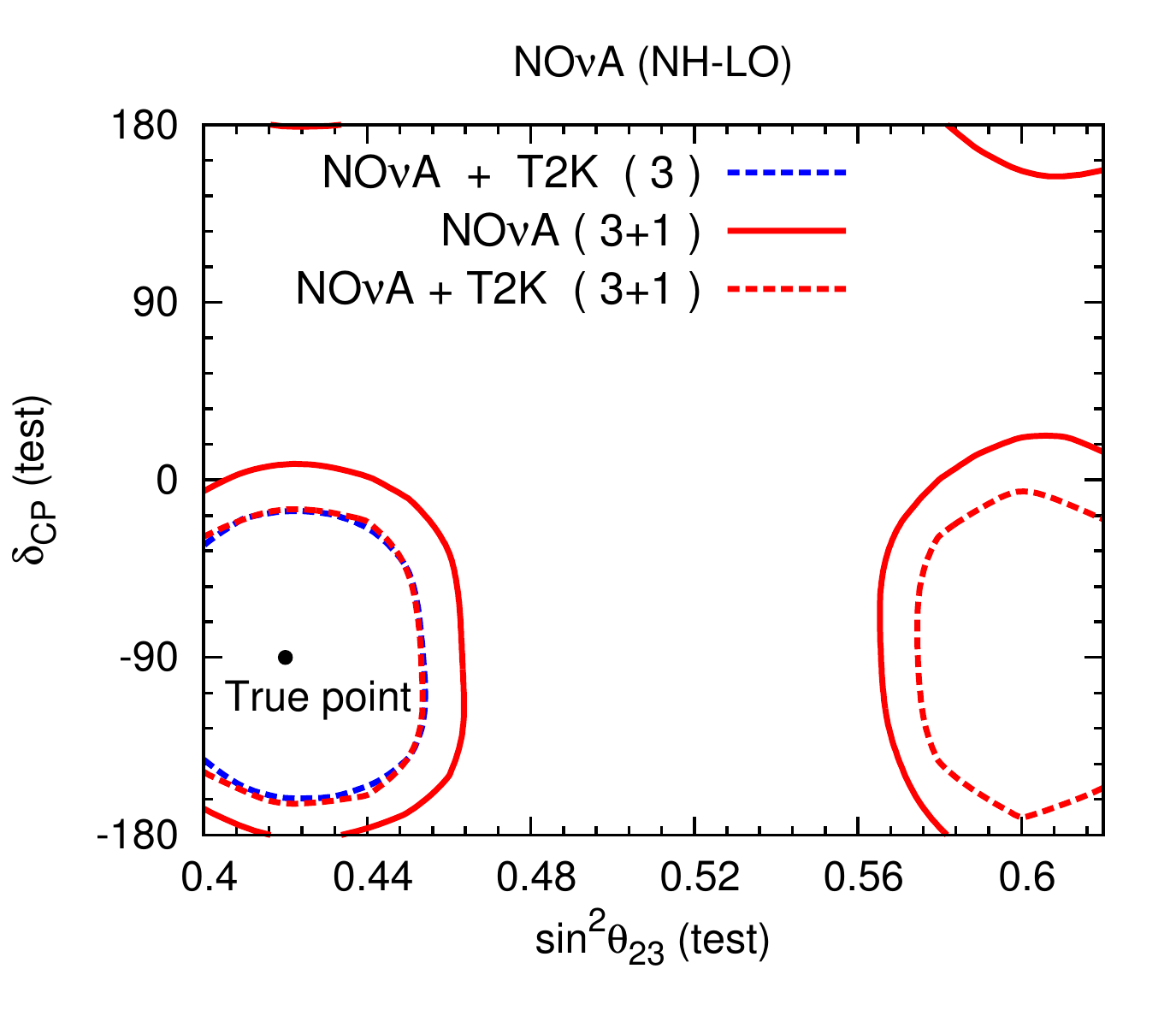}
\includegraphics[width=0.49\linewidth]{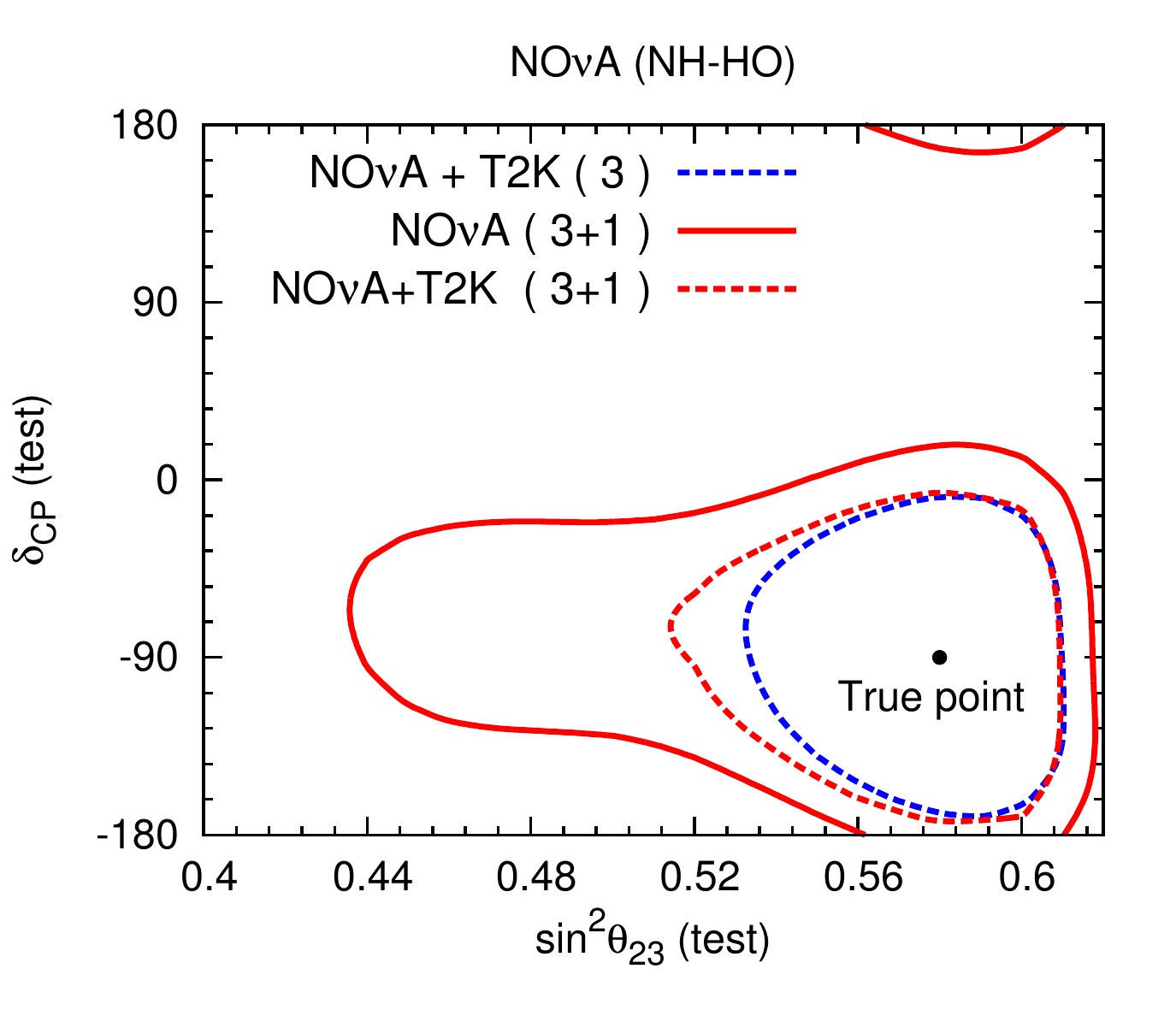}
 \caption{The allowed parameter space in $\theta_{23}-\delta_{\rm CP}$ plane for the long-baseline experiments T2K and NO$\nu$A. Oscillation parameter used for the analysis are given in  in Table \ref{table}. True value of $\sin ^2\theta_{23}$ for LO (HO) is considered as 0.42 (0.58). The true point is shown by the black dot in each plot.}
\label{parameterspace}
\end{center}
\end{figure}

Another way of representing these degeneracies among oscillation parameters is by using the bi-probability plot.  In this case, we calculate the oscillation probabilities for neutrino and anti-neutrino for a fixed  hierarchy-octant  combination for all possible  values of $\delta_{\rm CP}$ and display them in a neutrino-antineutrino probability plane  in  Fig. \ref{biprob}. The ellipses in the figure correspond to 3 flavor case, whereas the bands represent the oscillation probabilities in presence of sterile neutrino with all possible values of new phases $\delta_{14}$ and $\delta_{34}$. From the figure, it can be seen that the ellipses for LO and HO are very well separated for both hierarchies, whereas the ellipses for NH and IH for both LO and HO are overlapped with each other and give rise to degeneracies. Therefore,  NO$\nu$A experiment is more sensitive to octant of $\theta_{23}$ than that of mass hierarchy. While in $3+1$ paradigm, the bands are overlapped with each other for all combinations, which gives rise to new degeneracies. The additional degeneracies between lower and higher octants along with the standard ones, indicates that experiment is loosing its sensitivity in presence of sterile neutrino.\\

Next, we show the allowed parameter space in $\theta_{23}-\delta_{\rm CP}$ plane for each hierarchy-octant combination as given in Fig. \ref{parameterspace}.  In order to obtain the allowed parameter space, we simulate the true event spectrum with oscillation parameters given in Table \ref{table} and compare it with test event spectrum by varying test values of $\theta_{23}$,   $\delta_{\rm CP}$ in their allowed ranges and doing marginalization over $|\Delta m_{31}|^2$ for standard paradigm. In the $3+1$ case, we also do marginalization over  new phases $\delta_{14}$ and $\delta_{34}$.  The solid blue (red) curve in the figure is for standard paradigm (3+1 case) for NO$\nu$A experiment, whereas the dashed curves are for the combined analysis of T2K and NO$\nu$A experiments. The plots in the  left (right) panel correspond to  lower (higher) octant. From the top panel of the figure, it can be seen  that the allowed parameter space in the presence of sterile neutrino is enlarged which  indicates that the degeneracy resolution capability is deteriorated significantly. However, the synergy of T2K and  NO$\nu$A improves the degeneracy resolution capability.
\begin{figure}[!htb]
\begin{center}
\includegraphics[scale=0.49]{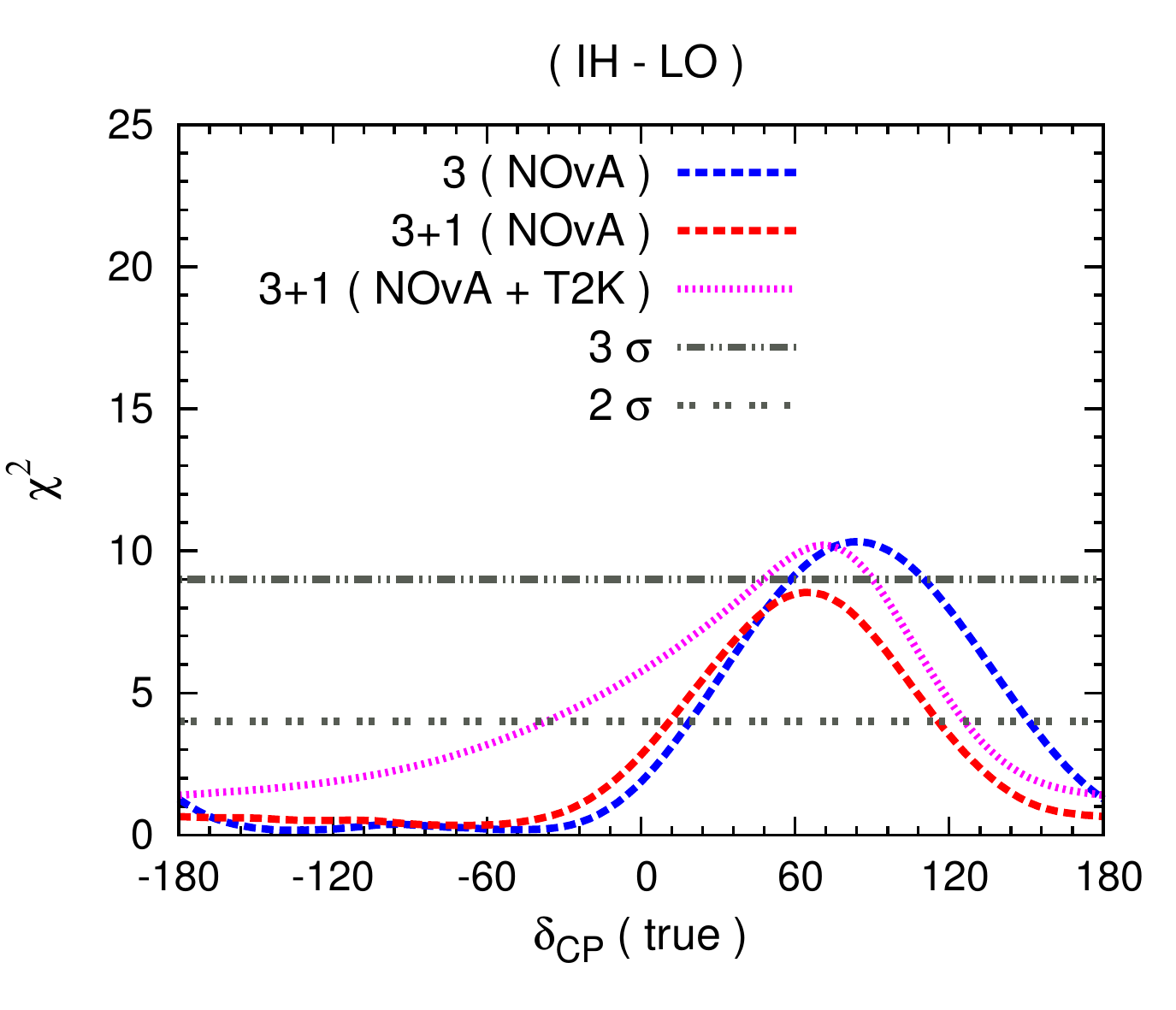}
\includegraphics[scale=0.49]{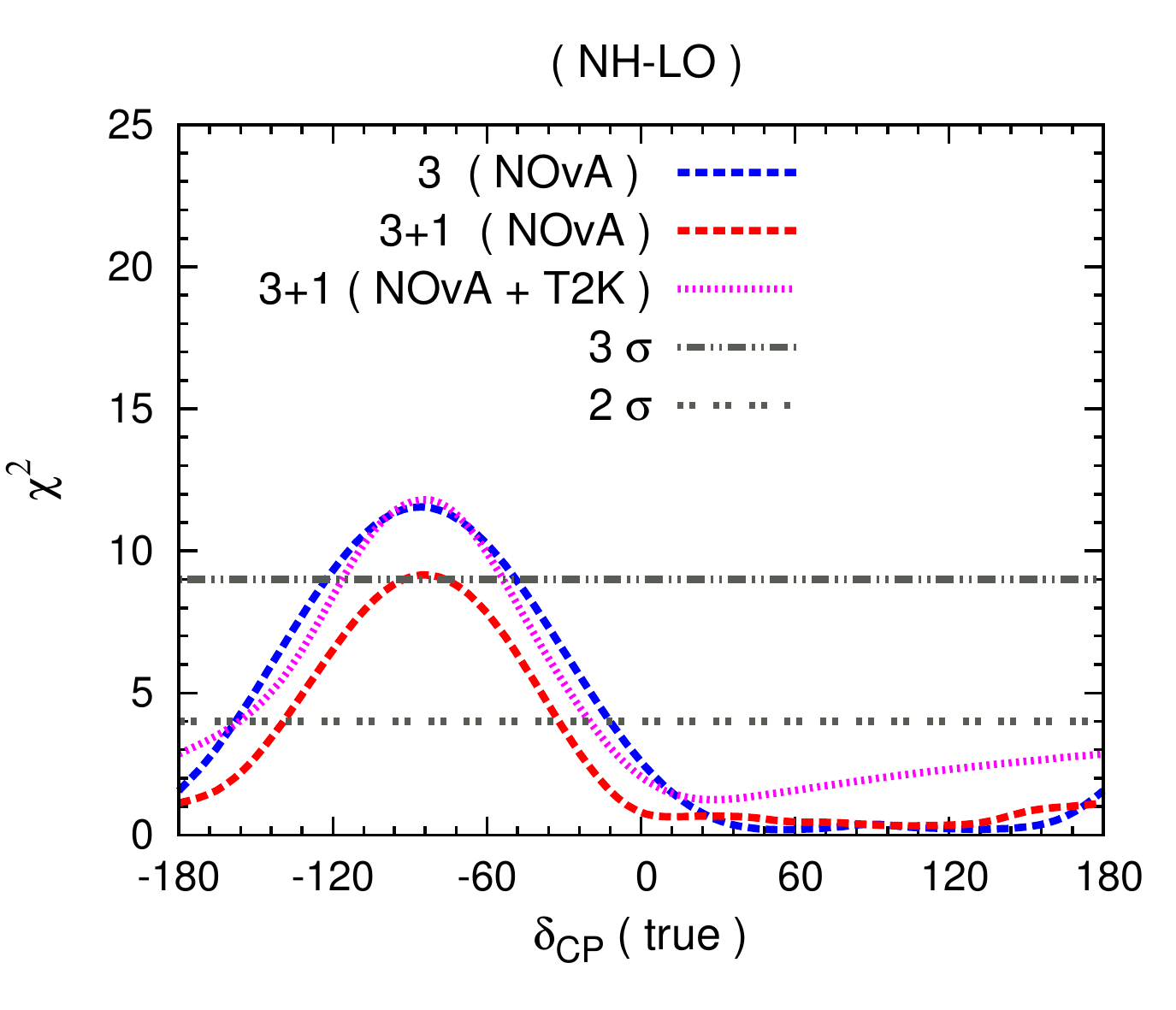} 
\includegraphics[scale=0.49]{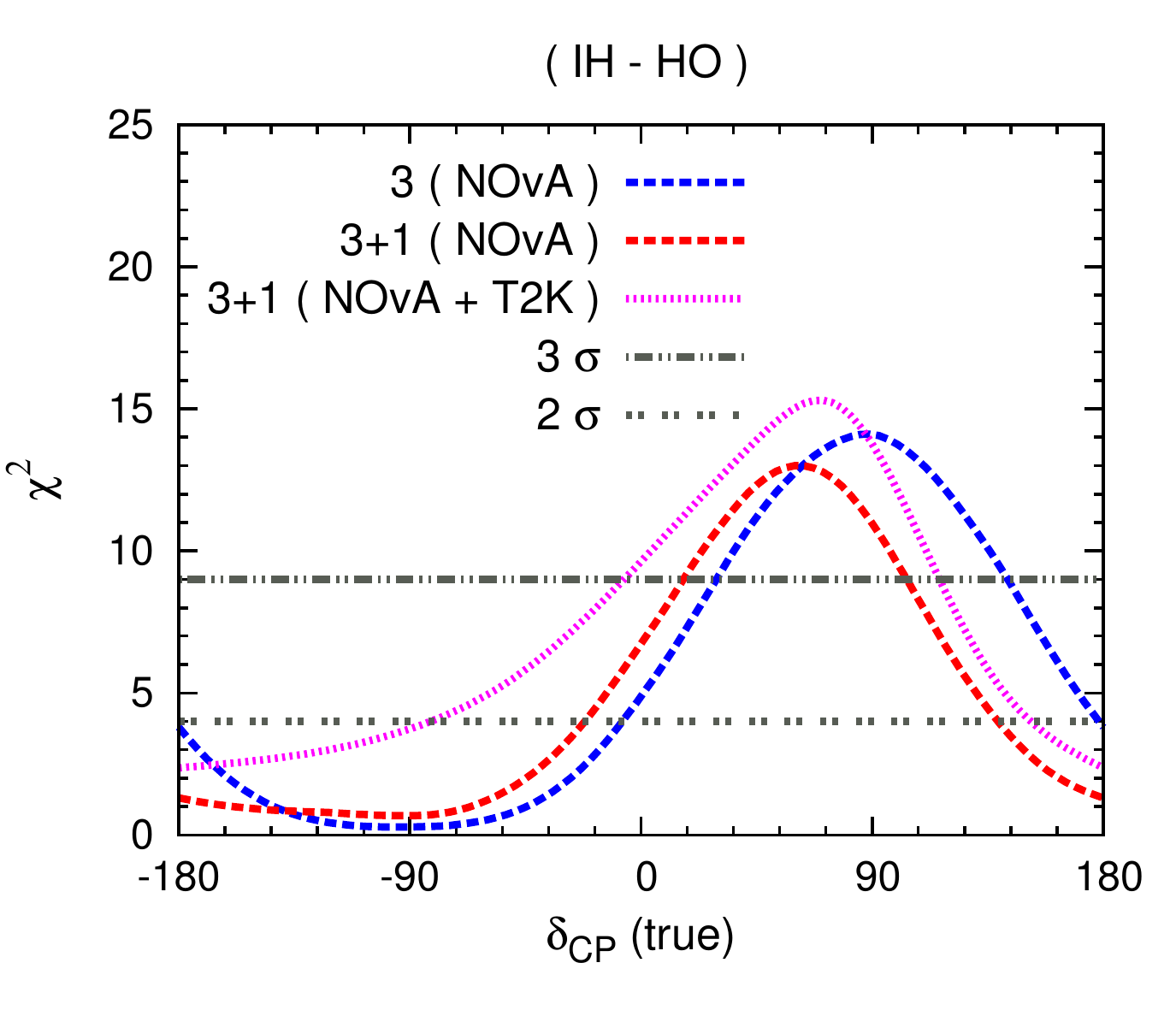}
\includegraphics[scale=0.49]{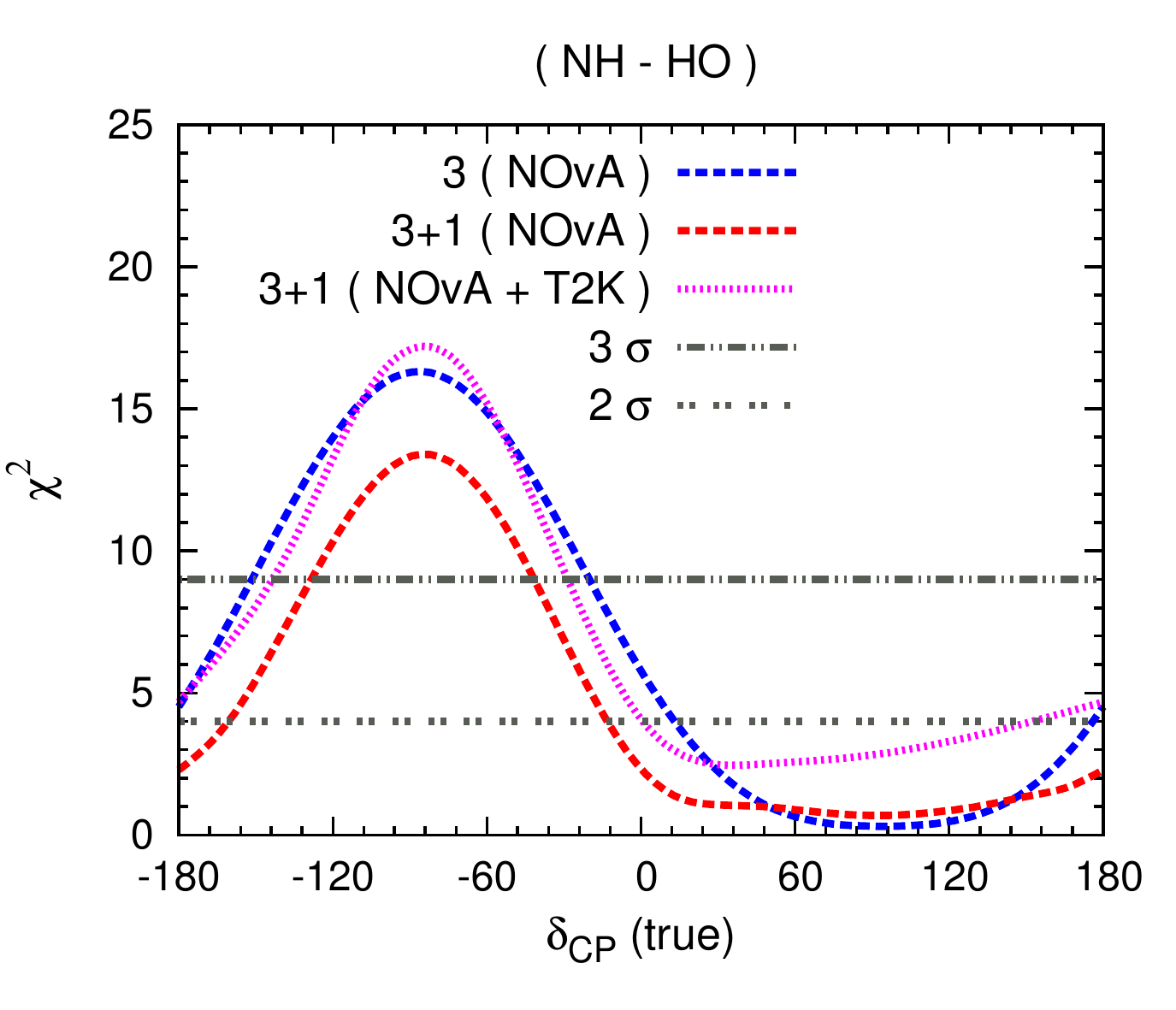}
\caption{MH sensitivity as a function of true values of $\delta_{\rm CP}$. The left (right) panel is for inverted (normal) hierarchy and the upper (bottom) panel is for LO (HO).}
\label{mh}
\end{center}
\end{figure}
\section{Mass Hierarchy Sensitivity}
In this section, we discuss how mass hierarchy sensitivity of NO$\nu$A experiment gets modified in presence of sterile neutrino. In order to obtain the MH sensitivity, we simulate the event spectrum by assuming true hierarchy as normal (inverted) and test hierarchy as inverted (normal). We obtain $\chi^2$ by comparing true and test event spectra as discussed in section III (Eqns.(7-9)). While doing the calculation, we do marginalisation over $\delta_{\rm CP},~\theta_{23},~|\Delta m^2_{31}|$ for standard paradigm and in addition to this, we also do marginalisation over $\delta_{14}$ and $\delta_{34}$ for (3+1) case, in their corresponding ranges as shown in 
Table \ref{table}. In Fig. \ref{mh}, we present the hierarchy determination sensitivity of NO$\nu$A. The left (right) panel corresponds to inverted (normal) hierarchy as true hierarchy, while lower (upper) panel corresponds to lower (higher) octant.  From the figure, one can see that the wrong mass hierarchy can be ruled out significantly above 2$\sigma$ in the favourable regions, i.e., lower half-plane (upper half-plane) for NH (IH) in the standard paradigm  as shown by dotted blue curves. Whereas, in presence of sterile neutrino the $\delta_{\rm CP}$ coverage for the mass hierarchy sensitivity is significantly reduced  as shown by the dotted red curves.
At the same time the combined analysis of T2K with NO$\nu$A  shows a significant increase in MH sensitivity due to increase of $\delta_{\rm CP}$ coverage as shown in Fig. \ref{mh} by magenta curves.
\section{Effect on maximal CP-violation exclusion sensitivity}
One of the main objectives of long-baseline experiments is to look for non-zero CP-violation in leptonic sector. Further, the recent global-fit data provide us the hint for maximal CP-violation with $\delta_{\mathrm{CP}}\approx-90^{\circ}$ \cite{deSalas:2017kay}. Therefore, in this section, we check the compatibility of observed data with the maximal CP-violation hypothesis in presence of sterile neutrino. In order to quantify our analysis, we define the parameter,
\begin{equation}
\Delta\chi^2_{\mathrm{MCP}} = \underset{\delta_{\mathrm{CP}}^{\mathrm{test}} = \{90^{\circ},-90^{\circ}\}}{\mathrm{ min}} \left[ \chi^2(\delta_{\mathrm{CP}}^{\mathrm{test}})-  \chi^2(\delta_{\mathrm{CP}}^{\mathrm{true}})\right].
\end{equation}
\begin{figure}
\includegraphics[scale=0.77]{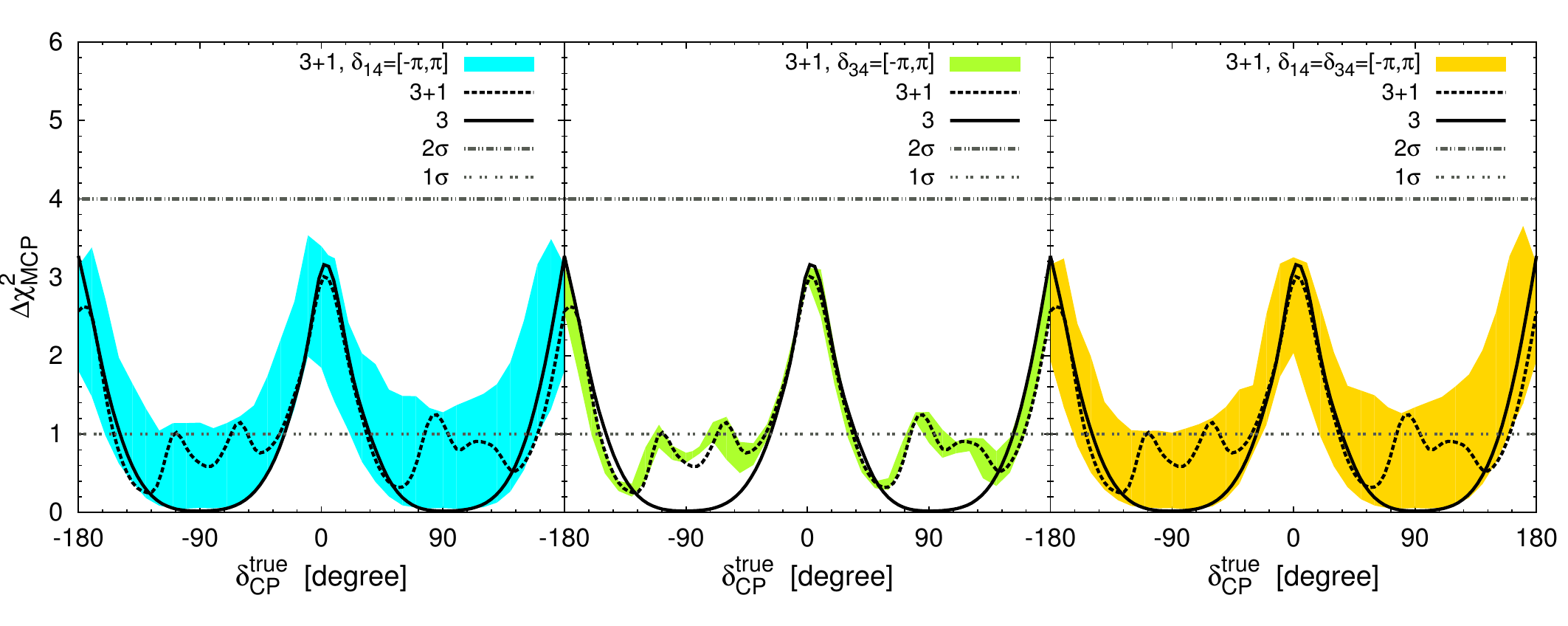}
\caption{Maximal CP-violation exclusion sensitivity as a function of true value of $\delta_{\mathrm{CP}}$ for NO$\nu$A experiment. We assume the true hierarchy as  NH and use the values of the oscillation parameters as given in  Table \ref{table}.}
\label{mcp} 
\end{figure}
We show the sensitivities of excluding the maximal CP-violation ($\Delta\chi^2_{\mathrm{MCP}}$) for NO$\nu$A as a function of true values of $\delta_{\mathrm{CP}}$   for both standard 3 flavor and 3+1 flavor paradigms in Fig. \ref{mcp}.
In each panel the black solid curve corresponds to the sensitivity in 3-neutrino scenario, while the black dashed  curve is for 3+1 paradigm with the additional new phases ($\delta_{14},\delta_{34}$) set to zero.
In the left (middle) panel, we show the effect of additional phase $\delta_{14}(\delta_{34})$ in 3+1 scenario, whereas in the right panel, we show the effect of both phases.
   While doing the analysis, we use the true values of oscillation parameters as given in the Table I.  For each true choice of  $\delta_{\mathrm{CP}}$ (=$\delta_{13}$),  we do marginalization over the phases: $\delta_{13},\delta_{14},\delta_{34},$  for $90^{\circ}$ and $-90^{\circ}$. In addition to this, we do marginalisation over the $\theta_{23}$ and $\Delta m^2_{31}$. Here, we assume the mass hierarchy to be normal. 
 It can be seen from the figure that, for 3 flavor scenario the maximal CP violation hypothesis can be excluded by $>1\sigma$ C.L.  for values of $\delta_{\mathrm{CP}}$ near to the region of 0, $\pm \pi$ for NO$\nu$A experiment. The dashed black curve corresponds to the 3+1 scenario  which is showing an oscillatory behaviour due to the additional mixing angles ($\theta_{14},\theta_{24},\theta_{34}$). From the figure, it should be noted that the CP-phase $\delta_{14}$ has large impact on the sensitivity compared to the $\delta_{34}$.  

\section{Implications  on Neutrinoless double beta decay} 
 In this section, we would like to see the implication of the eV scale sterile neutrino on some low-energy phenomena, like neutrinoless double beta decay ($0\nu \beta \beta$).  One of the important features of  $0 \nu \beta \beta$ process is that it violates the lepton number by two units   and hence, its experimental observation would not only ascertain the Majorana  nature of light neutrinos, but also can 
 provide the  absolute scale of lightest active neutrino mass. Various neutrinoless double beta decay experiments like KamLAND-Zen \cite{KamLAND-Zen:2016pfg}, GERDA \cite{Agostini:2013mzu}, EXO-200 \cite{Albert:2014awa} etc., have provided bounds on the half-life (${\cal T}_{1/2}$)   of this process on various isotopes, which can be translated as a bound on  effective Majorana mass parameter $|M_{ee}|$ \cite{Doi:1985dx,Haxton:1985am} as,
 \bea
 ({\cal T}_{1/2})^{-1}= Q \left | \frac{{\cal M}_\nu}{m_e}\right |^2 |M_{ee}|^2\;,
 \eea
 where $Q$ is the phase space factor, ${\cal M}_\nu$ is the nuclear matrix element (NME) and $m_e$ is the electron mass. Recently $0\nu \beta \beta$ experiments involving $^{\rm 76}{\rm Ge}$, GERDA \cite{Agostini:2013mzu}, $^{\rm 136}{\rm Xe}$ EXO-200 \cite{Albert:2014awa} have provided the upper limit on $|M_{ee}|$ as $\sim (0.2-0.4)$ eV, using the available results on nuclear matrix elements (NME) from literature.  The current best   upper  limit on  $|M_{ee}|$   has been reported by KamLAND-Zen Collaboration \cite{KamLAND-Zen:2016pfg} as $|M_{ee}| < (0.061-0.165)$ eV at 90\% CL. The next generation experiments are planning to probe towards   the effective mass range: $|M_{ee}| < (10^{-3}-10^{-2})$ eV  regime, and hopefully, they can cover the inverted mass  hierarchy region of parameter space. The impact of an eV-scale sterile neutrino on neutrinoless double-beta decays is studied in \cite{Huang:2019qvq} following the Bayesian statistical approach, where it has been shown that  a null signal from the future $0 \nu \beta \beta$ decay experiments with a sensitivity to $|M_{ee}| \approx {\cal O}(10^{-2})$ would be able to set stringent constraints on the mass of the sterile neutrino as well as the the active-sterile mixing angle.
  
\begin{figure}[!htb]
\begin{center}
\includegraphics[scale=0.55]{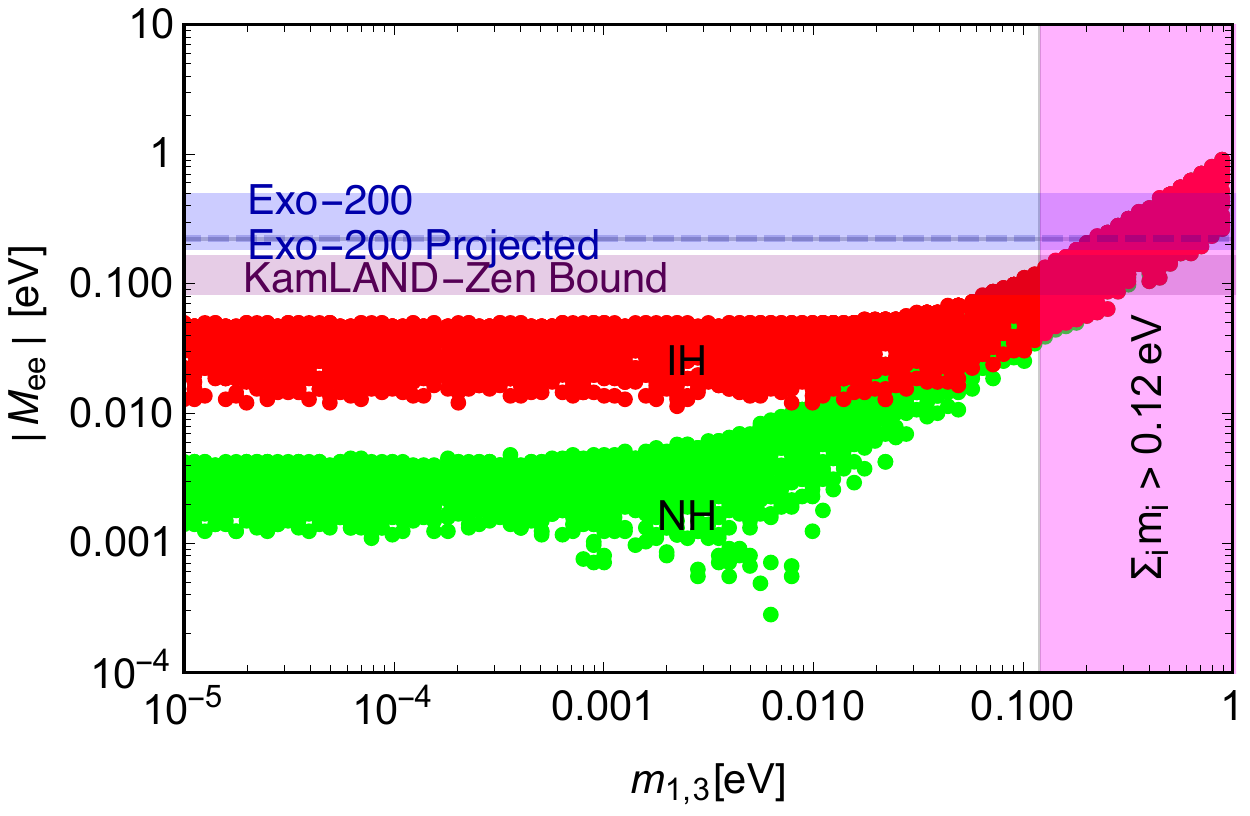}\\
\vspace*{0.2true in}
\includegraphics[scale=0.55]{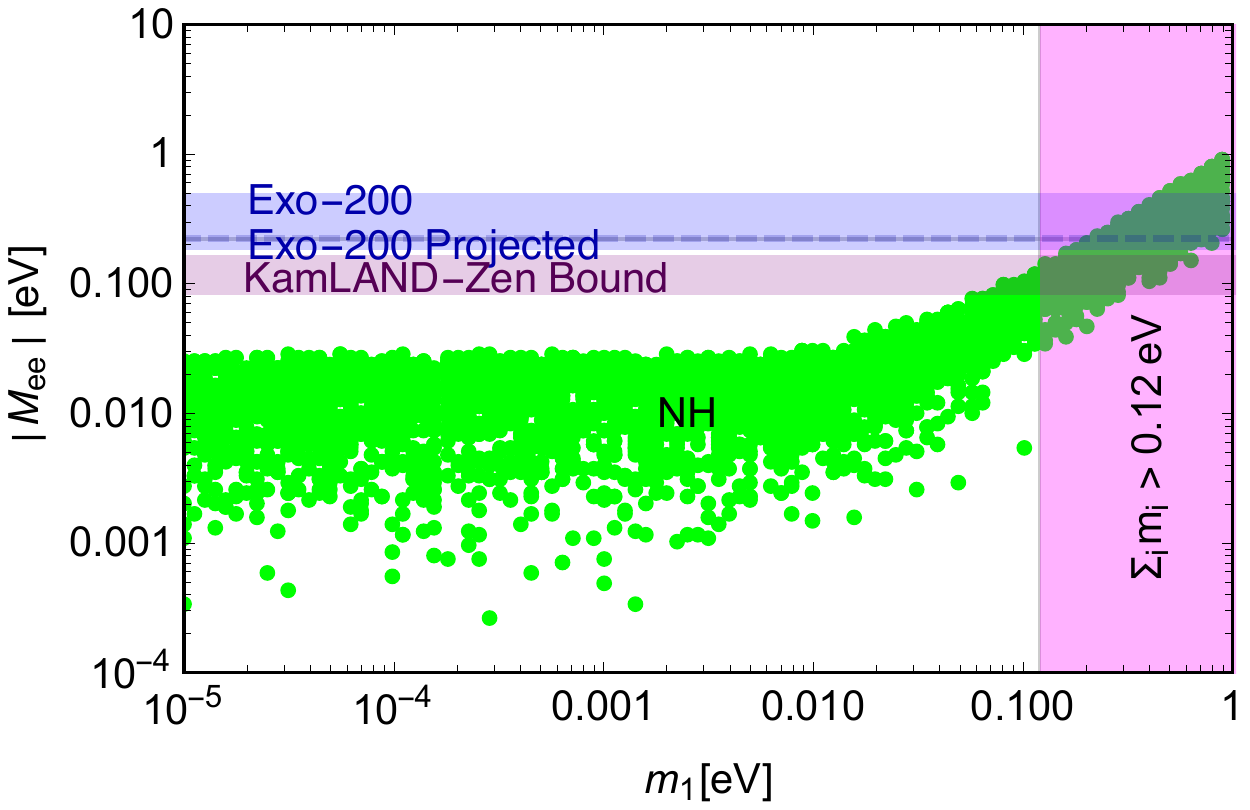} 
\hspace*{0.2true in}
\includegraphics[scale=0.55]{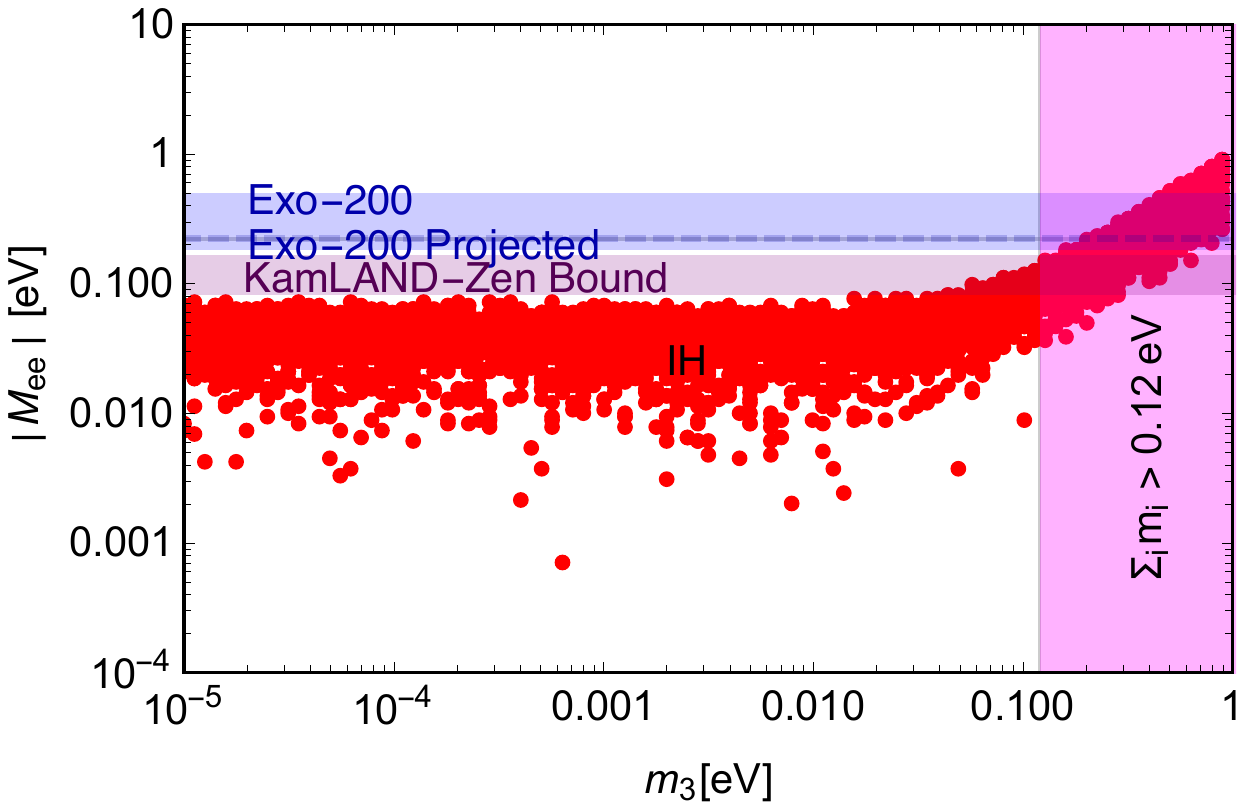}
\caption{Variation of the effective Majorana mass parameter $|M_{ee}|$ with the lightest neutrino mass, where the top panel is for the standard  three generation of neutrinos, and the bottom panels are due to the presence of an additional eV-scale neutrino. }\label{nbd}
\end{center}
\end{figure}

 The effective Majorana mass, which is  the key parameter of $0 \nu \beta \beta$ decay process   is defined in the standard three neutrino formalism as
\begin{subequations}
\begin{eqnarray}
\left| M_{ee} \right|=
  \bigg| U^2_{e1}\, m_1 + U^2_{e2}\, m_2 e^{i \alpha} + U^2_{e3}\, m_3 e^{i \beta} \bigg| \;, 
\label{eq:mee-std}
\end{eqnarray}
\end{subequations}
where $U_{e i}$ are the PMNS matrix elements and $\alpha$, $\beta$ are the Majorana phases. In terms of the lightest neutrino mass $m_l$ and the atmospheric and solar mass-squared differences, it can be expressed for NH and IH as
\begin{eqnarray}
\left| M_{ee} \right|_{\rm NH}=
  \bigg| U^2_{e1}\, m_l + U^2_{e2}\, \sqrt{\Delta m^2_{ \rm sol}+m_l^2} \;e^{i \alpha} + U^2_{e3}\; \sqrt{\Delta m^2_{\rm atm}+m_l^2}\,  e^{i \beta} \bigg|,
\end{eqnarray}
 and 
 \begin{eqnarray}
\left| M_{ee} \right|_{\rm IH}=
  \bigg| U^2_{e1}\, \sqrt{\Delta m^2_{\rm atm}-\Delta m^2_{\rm sol} +m_l^2} + U^2_{e2}\, \sqrt{\Delta m^2_{ \rm atm}+m_l^2} \;e^{i \alpha} + U^2_{e3}\; m_l\,  e^{i \beta} \bigg|.
\end{eqnarray}
Analogously, one can obtain the expression for $|M_{ee}|$ in the presence of an additional sterile neutrino as
\begin{eqnarray}
\left| M_{ee} \right|=
  \bigg| U^2_{e1}\, m_1 + U^2_{e2}\, m_2 e^{i \alpha} + U^2_{e3}\, m_3 e^{i \beta}+ U^2_{e4}\, m_4 e^{i \gamma} \bigg|.
\end{eqnarray}
Now varying the PMNS matrix elements as well as the Dirac CP phase within their $3\sigma$ range \cite{deSalas:2017kay} and the Majorana phases $\alpha$ and $\beta$ between $[0,2 \pi]$, we show the variation of $|M_{ee}|$ for three generation of neutrinos in the top panel of Fig. \ref{nbd}. Including the contributions from the eV scale sterile neutrino the corresponding plots are shown in the bottom panel, where the  left panel is for  NH and the right one for IH. In all these plots, the horizontal regions represent the bounds on effective Majorana mass from various $0 \nu \beta \beta$  experiments, while the vertical shaded regions are disfavoured from Planck data on the sum of light neutrinos, where the current bound is  $\Sigma_i m_i < 0.12$~eV from Planck+WP+highL+BAO 
data  at 95\% C.L. \cite{Ade:2015xua}. It should be noted that with the inclusion of an eV scale sterile neutrino,  part of the  the parameter space of $|M_{ee}|$ (for IH) is  within the  
sensitivity reach of KamLAND-Zen experiment. Furthermore, there is also some overlap regions between NH and IH cases. Thus, the future $0 \nu \beta \beta$ decay experiments may shed light on several issues related the nature of neutrinos.  

{\bf Comment on sensitivity reach of future experiments:}

\begin{figure}[!htb]
\includegraphics[scale=0.6]{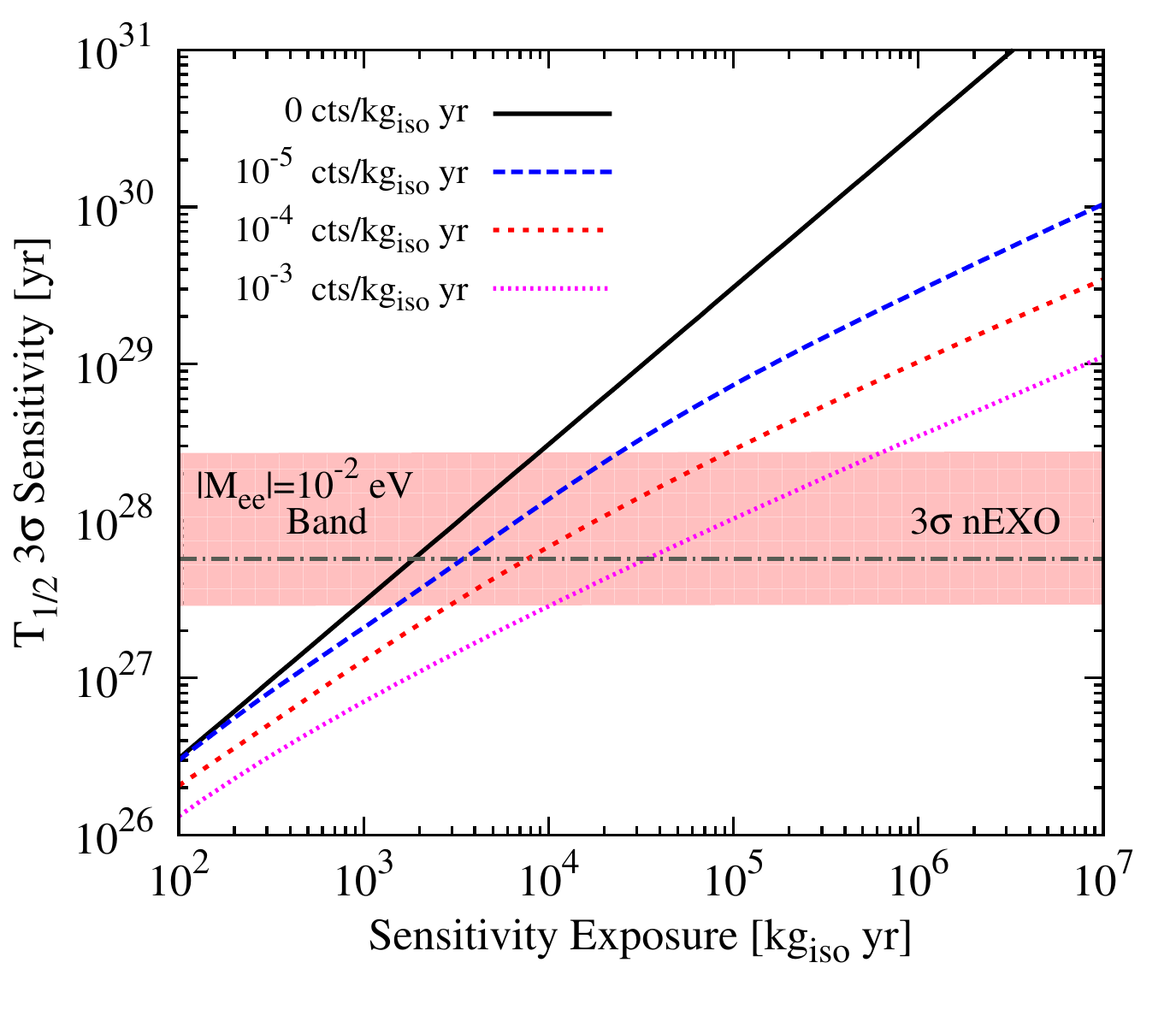}
\caption{$^{136}Xe$ discovery sensitivity as a function of sensitivity exposure for a representative set  of sensitive background levels. The  black,  blue, red and magenta lines correspond to  the values of  sensitive background levels of $0, 10^{-5},  10^{-4}$ and $10^{-3}$  ${\rm cts}/({\rm kg_{iso} yr})$ respectively.}
\label{beta}
\end{figure}

Here, we present a brief discussion on the sensitivity of eV-scale sterile neutrino in the future
$^{136}Xe$ experiment. The discovery sensitivity of an experiment is characterized by the value of half-life $({\cal T}_{1/2}$) for which it has 50\%  probability of measuring a $3 \sigma$ signal, above the background, defined as \cite{Agostini:2017jim, N.:2019cot}
\bea
{\cal T}_{1/2}=\frac{\ln 2 N_A \varepsilon}{m_a S_{3\sigma}(B)}\;,\label{t1/2}
\eea
where $N_A$ is the Avogadro's number, $m_a$ denotes the atomic mass of the $Xe$ isotope, $B=\beta \varepsilon$ ($\beta$ and $\varepsilon$ stand for the background and exposure sensitivity),  and $S_{3 \sigma}$ signifies the value for which 50\% of the measurements would give a signal above $B$, which can be calculated  assuming a Poisson distribution 
\bea
1-CDF_{\rm Poisson}(C_{3 \sigma}|S_{3 \sigma}+B)=50 \%.\label{sens}
\eea
Here $C_{3 \sigma}$ indicates the number of counts for which $CDF_{\rm Poisson}(C_{3 \sigma}|B)=3 \sigma$ and the continuous Poisson distribution  can be defined in terms of incomplete gamma function as
\bea
CDF_{\rm Poisson}(C|\mu)= \frac{\Gamma(C+1,\mu)}{\Gamma(C+1)}\;.\label{cdf}
\eea
Thus, with Eqns. (\ref{t1/2}) and (\ref{cdf}),  we show in Fig. \ref{beta}, the discovery sensitivity of ${\cal T}_{1/2}$ for $^{136}Xe$  as a function of $\varepsilon$ for various values of $\beta$.  The red band corresponds to a representative value  of $|M_{ee}| = 10^{-2}$ eV in the presence of a sterile neutrino (expressed in terms of the half-life ${\cal T}_{1/2}$ using Eqn. (16)),  and varying  the   parameters in the PMNS matrix within their $3 \sigma$ allowed ranges and also taking into account the uncertainty in  the nuclear matrix element (${\cal M}_\nu$).
 In Fig. \ref{beta}, the dotted black line represents the future 3$\sigma$ sensitivity of nEXO
 \cite{Kharusi:2018eqi}, which is ${\cal T}_{1/2} = 5.7 \times 10^{27}$ years. The black, blue, red, and magenta lines correspond to different values of the sensitive background levels of 0, $10^{-5}, 10^{-4}$ and $10^{-3} ~{\rm cts}/({\rm kg_{iso} yr})$ respectively.   From the figure, we can see that for a sensitive background level of $10^{-4}~ {\rm cts}/({\rm kg_{iso}yr})$,  the   $10^{-2}$ eV region could be probed with a sensitive exposure of $\sim 10^4~{\rm  kg_{iso}yr}$.

\section{Conclusion}
The various short baseline anomalies hint towards existence of an eV scale sterile neutrino. If such neutrino exists, it can mix with active neutrinos and affect the sensitivities of long-baseline experiments.  
As one of the main objectives of currently running long-baseline experiments is to determine mass hierarchy of neutrinos, in this paper,  we discussed the effect of active-sterile mixing on the  degeneracy resolution capability  and MH sensitivity of NO$\nu$A experiment. We found that introduction of sterile neutrino gives rise to new kind of degeneracies among the oscillation parameters which  results in reduction of $\delta_{\rm CP}$ coverage for MH sensitivity of NO$\nu$A experiment.  We also found that addition of T2K data helps in resolving the degeneracies among the oscillation parameters  and for MH sensitivity analysis, results  a significant increase in $\delta_{\rm CP}$ coverage for one additional sterile neutrino. 
We further scrutinized the  compatibility of the observed data with the maximal CP-violation hypothesis in presence of sterile neutrino.
We have also studied the effect sterile neutrino on neutrinoless double beta decay process and shown that  the inclusion of an eV scale sterile neutrino can enhance the value  of the effective mass parameter $|M_{ee}|$,  and for IH it could  be    within the  
sensitivity reach of KamLAND-Zen experiment. We also comment on the sensitivity reach of  future $^{136}Xe$ experiments for exploring the presence of  eV-scale sterile neutrino
and found that  for a sensitive background level of $10^{-4}~ {\rm cts}/({\rm kg_{iso}yr})$,  the   $10^{-2}$ eV region of  effective Majorana mass parameter ($|M_{ee}|$) could be probed with a sensitive exposure of $\sim 10^4~{\rm  kg_{iso}yr}$. \\

{\bf Acknowledgements} One of the authors (Rudra Majhi) would like to thank Department of Science \& Technology (DST) Innovation   in Science Pursuit for Inspired Research (INSPIRE) for financial support. The work of RM is supported by SERB, Govt. of India through grant no. EMR/2017/001448.  We sincerely thank Dinesh K. Singha for his timely help related to the computational work. Also thanks to Akshaya Chatla for discussion regarding GLoBES. We gratefully acknowledge the use of CMSD HPC  facility of Univ. of Hyderabad to carry out computations in this work.


\medskip
\begin{thebibliography}{60}
\bibitem{Tanabashi:2018oca} M. Tanabashi et al. (Particle Data Group), Phys. Rev. D{\bf 98}, 030001 (2018).

\bibitem{Agarwalla:2014fva} S. K. Agarwalla, Adv. High Energy Phys. {\bf 2014}, 457803 (2014), 1401.4705.

\bibitem{Feldman:2013vca} G. J. Feldman, J. Hartnell, and T. Kobayashi, Adv. High Energy Phys. {\bf 2013}, 475749 (2013),
1210.1778.
 
\bibitem{Barger:2001yr} V. Barger, D. Marfatia, and K. Whisnant, Phys. Rev. D{\bf 65}, 073023 (2002), hep-ph/0112119.
 
 \bibitem{Aguilar:2001ty} A. Aguilar-Arevalo et al. (LSND), Phys. Rev. D{\bf 64}, 112007 (2001), hep-ex/0104049.
 
 \bibitem{Aguilar-Arevalo:2013pmq} A. A. Aguilar-Arevalo et al. (MiniBooNE), Phys. Rev. Lett. {\bf 110}, 161801 (2013), 1303.2588.
 
\bibitem{Lasserre:2012vy} T. Lasserre, Nucl. Phys. Proc. Suppl. {\bf 235-236}, 214 (2013), 1209.5090.

\bibitem{Mention:2011rk} G. Mention, M. Fechner, T. Lasserre, T. A. Mueller, D. Lhuillier, M. Cribier, and A. Letourneau, Phys. Rev. D{\bf 83}, 073006 (2011), 1101.2755.

\bibitem{Ko:2016owz} Y. J. Ko et al. (NEOS), Phys. Rev. Lett. {\bf 118}, 121802 (2017), 1610.05134.

\bibitem{Alekseev:2018efk} I. Alekseev et al. (DANSS), Phys. Lett. B{\bf 787}, 56 (2018), 1804.04046.
 
 \bibitem{Ashenfelter:2018iov} J. Ashenfelter et al. (PROSPECT), Phys. Rev. Lett. {\bf 121}, 251802 (2018), 1806.02784.
 
 \bibitem{Lhuillier:2015fga} D. Lhuillier, AIP Conf. Proc. {\bf 1666}, 180003 (2015).
 
 \bibitem{Anselmann:1994ar} P. Anselmann et al. (GALLEX), Phys. Lett. B{\bf 342}, 440 (1995).

 \bibitem{Hampel:1997fc} W. Hampel et al. (GALLEX), Phys. Lett. B{\bf 420}, 114 (1998).
 \bibitem{Kaether:2010ag} F. Kaether, W. Hampel, G. Heusser, J. Kiko, and T. Kirsten, Phys. Lett. B{\bf 685}, 47 (2010),
1001.2731.
 \bibitem{Abdurashitov:1996dp} D. Abdurashitov et al., Phys. Rev. Lett. 77, 4708 (1996).
 \bibitem{Giunti:2012tn} C. Giunti, M. Laveder, Y. F. Li, Q. Y. Liu, and H. W. Long, Phys. Rev. D{\bf 86}, 113014 (2012), 1210.5715.

 \bibitem{Giunti:2010zu} C. Giunti and M. Laveder, Phys. Rev. C{\bf 83}, 065504 (2011), 1006.3244.

\bibitem{Abazajian:2012ys} K. N. Abazajian et al. (2012), 1204.5379.

\bibitem{Abdurashitov:2009tn} J. N. Abdurashitov et al. (SAGE), Phys. Rev. C{\bf 80}, 015807 (2009), 0901.2200.
 
 \bibitem{Aguilar-Arevalo:2018gpe} A. A. Aguilar-Arevalo et al. (MiniBooNE), Phys. Rev. Lett. {\bf 121}, 221801 (2018), 1805.12028.
 
 \bibitem{Adamson:2017uda} P. Adamson et al. (MINOS+), Phys. Rev. Lett. {\bf 122}, 091803 (2019), 1710.06488.
 
 \bibitem{NOvA:2018gge} M. A. Acero et al. (NOvA), Phys. Rev. D{\bf 98}, 032012 (2018), 1806.00096.

\bibitem{Dutta:2016glq} D. Dutta, R. Gandhi, B. Kayser, M. Masud, and S. Prakash, JHEP {\bf 11},122(2016),1607.02152.

\bibitem{Ska:2018} S. K. Agarwalla, S. S. Chatterjee, and A. Palazzo, JHEP {\bf 04}, 091 (2018), 1801.04855.

\bibitem{Gupta:2018qsv} S. Gupta, Z. M. Matthews, P. Sharma, and A. G. Williams, Phys. Rev. D{\bf 98}, 035042 (2018), 1804.03361.

\bibitem{Ghosh:2017atj} M. Ghosh, S. Gupta, Z. M. Matthews, P. Sharma, and A. G. Williams, Phys. Rev. D{\bf 96}, 075018 (2017), 1704.04771.

\bibitem{Agarwalla:2016mrc} S. K. Agarwalla, S. S. Chatterjee, A. Dasgupta, and A. Palazzo, JHEP {\bf 02}, 111 (2016), 1601.05995.

\bibitem{Palazzo:2015gja} A. Palazzo, Phys. Lett. B{\bf 757}, 142 (2016), 1509.03148.

\bibitem{Dewhurst:2015aba} D. Dewhurst (T2K), in {\it Proceedings, Topical Research Meeting on Prospects in Neutrino Physics (NuPhys2014): London, UK, December 15-17, 2014} (2015), 1504.08237.

\bibitem{Bhattacharya:2011ee} B. Bhattacharya, A. M. Thalapillil, and C. E. M. Wagner, Phys. Rev. D{\bf 85}, 073004 (2012), 1111.4225.

\bibitem{Chatla:2018sos} A. Chatla, S. Rudrabhatla, and B. A. Bambah, Adv. High Energy Phys. 2018, {\bf 2547358} (2018),1804.02818.

\bibitem{Choubey:2017cba} S. Choubey, D. Dutta, and D. Pramanik, Phys. Rev. D{\bf 96}, 056026 (2017), 1704.07269.

\bibitem{Agarwalla:2016xxa} S. K. Agarwalla, S. S. Chatterjee, and A. Palazzo, JHEP {\bf 09}, 016 (2016), 1603.03759.

\bibitem{Berryman:2015nua} J. M. Berryman, A. de Gouva, K. J. Kelly, and A. Kobach, Phys. Rev. D{\bf 92}, 073012 (2015), 1507.03986.

\bibitem{Choubey:2017ppj} S. Choubey, D. Dutta, and D. Pramanik, Eur. Phys. J. C{\bf 78}, 339 (2018), 1711.07464.

\bibitem{Gandhi:2015xza} R. Gandhi, B. Kayser, M. Masud, and S. Prakash, JHEP {\bf 11}, 039 (2015), 1508.06275.

\bibitem{deSalas:2017kay} P. F. de Salas, D. V. Forero, C. A. Ternes, M. Tortola, and J. W. F. Valle, Phys. Lett. B{\bf 782}, 633 (2018), 1708.01186.

\bibitem{Majhi:2019exu} R. Majhi, C. Soumya, and R. Mohanta, Springer Proc. Phys. {\bf 234}, 341 (2019).

\bibitem{Li:2018ezt} W. Li, J. Ling, F. Xu, and B. Yue, JHEP {\bf 10}, 021 (2018), 1808.03985.

\bibitem{Esteban:2018azc} I. Esteban, M. C. Gonzalez-Garcia, A. Hernandez-Cabezudo, M. Maltoni, and T. Schwetz, JHEP {\bf 01}, 106 (2019), 1811.05487.

\bibitem{Gariazzo:2017fdh} S. Gariazzo, C. Giunti, M. Laveder, and Y. F. Li, JHEP {\bf 06}, 135 (2017), 1703.00860.

\bibitem{Huber:2004gg} P. Huber, M. Lindner, and W. Winter, JHEP {\bf 05}, 020 (2005), hep-ph/0412199.

\bibitem{Huber:2009cw} P. Huber, M. Lindner, T. Schwetz, and W. Winter, JHEP {\bf 11}, 044 (2009), 0907.1896.

\bibitem{C.:2014ika} Soumya C, K. N. Deepthi, and R. Mohanta, Adv. High Energy Phys. 2016, {\bf 9139402} (2016), 1408.6071.

\bibitem{Abe:2017uxa} K. Abe et al. (T2K), Phys. Rev. Lett. {\bf 118}, 151801 (2017), 1701.00432.

\bibitem{Abe:2014tzr} K. Abe et al. (T2K), PTEP {\bf 2015}, 043C01 (2015), 1409.7469.

\bibitem{KamLAND-Zen:2016pfg} A. Gando et al. (KamLAND-Zen), Phys. Rev. Lett. {\bf 117}, 082503 (2016), [Addendum: Phys.
Rev. Lett. {\bf 117}, no.10, 109903 (2016)], 1605.02889.

\bibitem{Agostini:2013mzu} M. Agostini et al. (GERDA), Phys. Rev. Lett. {\bf 111}, 122503 (2013), 1307.4720.

\bibitem{Albert:2014awa} J. B. Albert et al. (EXO-200), Nature {\bf 510}, 229 (2014), 1402.6956.

\bibitem{Doi:1985dx} M. Doi, T. Kotani, and E. Takasugi, Prog. Theor. Phys. Suppl. {\bf 83}, 1 (1985).

\bibitem{Haxton:1985am} W. C. Haxton and G. J. Stephenson, Prog. Part. Nucl. Phys. {\bf 12}, 409 (1984).

\bibitem{Huang:2019qvq} G.-Y. Huang and S. Zhou, Nucl. Phys. B{\bf 945}, 114691 (2019), 1902.03839.

\bibitem{Ade:2015xua} P. A. R. Ade et al. (Planck), Astron. Astrophys. {\bf 594}, A13 (2016), 1502.01589.

\bibitem{Agostini:2017jim} M. Agostini, G. Benato, and J. Detwiler, Phys. Rev. D{\bf 96}, 053001 (2017), 1705.02996.

\bibitem{N.:2019cot} K. N. Vishnudath, S. Choubey, and S. Goswami, Phys. Rev. D{\bf 99}, 095038 (2019), 1901.04313.
 
\bibitem{Kharusi:2018eqi} S. A. Kharusi et al. (nEXO) (2018), 1805.11142.
\end{thebibliography}
 \end{document}